\documentclass[prb,twocolumn,floats,showpacs]{revtex4}
\usepackage{epsf}
\usepackage{graphicx}
\usepackage{epsfig}
\usepackage{psfrag}
\usepackage{amssymb}
\newcommand{\ket}{\rangle}
\newcommand{\bra}{\langle}

\newcommand{\up}{\uparrow}
\newcommand{\dow}{\downarrow}

\begin{document}

\title{Rotating spin-1 bosons in the lowest Landau level}

\author{J.W.~Reijnders, F.J.M.~van Lankvelt and K.~Schoutens}
\affiliation{Institute for Theoretical Physics, University of Amsterdam,
Valckenierstraat 65, 1018 XE~~Amsterdam}
\author{N.~Read}
\affiliation{Department of Physics, Yale University, P.O. Box 208120,
New Haven, CT 06520-8120}



\date{July 11, 2003}

\begin{abstract}

\noindent We present results for the ground states of a system of
spin-1 bosons in a rotating trap.  We focus on the dilute, weakly
interacting regime, and restrict the bosons to the quantum states
in the lowest Landau level (LLL) in the plane (disc), sphere or
torus geometries.  We map out parts of the zero temperature phase
diagram, using both exact quantum ground states and LLL mean field
configurations. For the case of a spin-independent interaction we
present exact quantum ground states at angular momentum $L\leq N$.
For general values of the interaction parameters, we present mean
field studies of general ground states at slow rotation  and of
lattices of vortices and skyrmions at higher rotation rates.
Finally, we discuss quantum Hall liquid states at ultra-high rotation.

\end{abstract}

\maketitle
\vspace{0.1in}
\section{Introduction}\label{I}
In recent years, there has been considerable progress in the art
of manipulating cold atoms.  By varying experimental conditions, a
number of quantum states of atomic matter have been realized.  In
these developments, an exciting theme is the parallel between
these states of atomic matter and states of electronic or vortex
matter that have been studied before.  Two important recent
advances are the study of {\em spin-full bosons in optical
traps\/} and the analysis of {\em bosons in rotating traps\/}.

Optical traps liberate the (hyperfine) spin degree of freedom of
spin-full atoms, as there is no polarizing magnetic field.  This
allows a variety of new phenomena, such as skyrmions, monopoles
and $\pi$-disclinations.  There are two different regimes,
depending on the sign of the spin-dependent interaction $c_2$\@.
If this interaction is repulsive ($c_2>0$), the system tends to
minimize the total spin and we speak of the anti-ferromagnetic or
``polar'' regime.  If, on the other hand, the interaction is
attractive ($c_2<0$) the system will tend to maximize the total
spin.  This is the ferromagnetic regime.  Examples of such systems
are spin-1 Bose-Einstein condensates (BEC) which can be realized
by trapping atoms such as $^{87}$Rb ($c_2<0$)\cite{myatt,barrett}
and $^{23}$Na ($c_2>0$)\cite{sta-k}\@.  In both cases, the ratio
$\gamma=c_2/c_0$ of the spin-dependent ($c_2$) and the
spin-independent ($c_0$) parts of the (contact) interaction is
small, typically a few percent, $|\gamma| \approx  0.01-0.05$.

Rotating the bosons within the trap leads to the formation of
quantum ground states with a certain amount of vorticity stored in
the system. Upon increasing the rotation rate, a single-component
(scalar) condensate with a repulsive interaction goes through the
following stages: (i) the nucleation of a single vortex, (ii) the
formation of a triangular (Abrikosov) lattice of vortices.
Theoretical analysis predicts that this will be followed by (iii),
a quantum melting of the lattice and the formation of a (series
of) quantum liquids, where the vorticity is spread uniformly over
the system. In the presence of internal degrees of freedom, such
as those associated with the spin-states of spin-1 atoms, a
similar sequence of quantum ground states is expected, with
additional structure provided by the internal (spin) degrees of
freedom, and by the presence of additional parameters such as
$\gamma$.

In this paper we study the ground states of spin-1 atoms in a
rotating harmonic trap, focusing on situations where a truncation
to the lowest Landau level (LLL) can be justified.  We consider
generic values of $\gamma$ in the repulsive regime $c_0>0$, and
also focus on the special case $\gamma=0$, where the interaction
has an $SU(3)$ symmetry \cite{reijnders}. Before we come to the
details of our analysis, we briefly summarize some of the results
in the literature for rotating bosons, with either a single
component (scalar case) or several components, such as spin 1
(vector case). Much of this work is based on the restriction to
the LLL.

\subsection{Non-rotating condensates and topological excitations}

The case of vector BEC without rotation has been investigated,
most often by mean-field theory using the spin-1 Gross-Pitaevskii
equations, or with the further approximation of neglecting the
kinetic energy term (the Thomas-Fermi approximation). There are
two regimes \cite{Ho}, as mentioned already. In the ferromagnetic
regime $c_2<0$, the ground state has maximum possible spin, $S=N$.
Such a spin state can be constructed by condensing all the bosons
into a single-particle state with $S_z=+1$, or as a spin-rotation
of this. In the opposite antiferromagnetic regime, $c_2>0$, the
ground state has minimal spin. Such a spin state can be
constructed by condensing all the bosons in the same $S_z=0$
single-particle spin state, or by taking a spin-rotation of this
(notice that this is distinct from all of the ferromagnetic
states). This ``polar'' state breaks spin-rotation symmetry, even
though the expectation value of the total spin, or of the spin
density, is very close to zero. The distinct ordered states, that
can be mapped onto each other by the (broken) symmetries of spin
rotation and phase rotation, are labelled by points in an order
parameter manifold (or target space). For the ferromagnetic case,
this manifold is $SO(3)$ \cite{Ho}, while for the
antiferromagnetic or polar case it is $S^1\times S^2/{\mathbb
Z}_2$ \cite{Ho,zhou}. These ordered states possess excitations
that can be described as topological defects in the order, either
with a singularity at a point surrounded by a ``core'', or without
a singularity at all. We describe these in more detail in the
Appendix, but those that carry non-zero vorticity are relevant to
the rotating case which we discuss next.

\subsection{Slow rotation: vortices and skyrmions}

If the number of bosons in a rotating trap is sufficiently large,
the effect of slow rotation can be studied in a mean field
framework.  For a single species of bosons with repulsive
interactions, the rotation is accommodated through the creation of
singular vortices, with vanishing particle density at the vortex
cores.  As the rotation rate is increased from zero, there is a
critical frequency at which a single vortex first appears in the
system, followed by additional vortices at still higher rotation.
With a spin degree of freedom, the system has several components
in which to store the angular momentum. There is the possibility
that a vortex core for one spin component is filled by another
spin component, leading to core-less vortices or ``skyrmions''. In
such configurations, the total particle density is nowhere zero,
and there is a smooth spin texture.  Mean field states of this
type for rotating spin-1 bosons have been investigated
theoretically by solving the spin-full Gross-Pitaevskii (GP)
equations \cite{yip,isoka,merminho,mimaki}. For attractive
interactions, in contrast, the BEC remains in a compact blob,
without any vortices, all the way up to the maximum rotation
frequency (the trap frequency).

\subsection{Lattices of vortices and skyrmions}

When several vortices are present in a rotating scalar boson
condensate with repulsive interactions, they line up in a
triangular (Abrikosov) vortex lattice
\cite{vortex-exp,vortex-the}\@. In a vector BEC one expects to
find similar lattices, built from the coreless vortices just
described. The details of all this depend crucially on the
relative strength $\gamma=c_2/c_0$ of the spin-dependent
interaction.  For $\gamma=0$, where the $SU(3)$ symmetry between
the different spin components is not broken, the lattice that is
expected upon rotation is composed of three intertwined triangular
lattices. The vortex cores do not overlap, so that the density is
(almost) uniform. This lattice has been shown to be independent of
the strength of the interaction by Kita {\it et
al.\/}\cite{mfskyr}\@. The vortex lattice shows a rich phase
diagram, however, when the interaction is spin-dependent.  For a
range of positive values of $\gamma$, a square lattice
composed of $\pi$-disclinations has been predicted \cite{mfskyr}.

\subsection{Scalar boson quantum liquids}
As the rotation increases, quantum fluctuations become more and
more important and beyond a critical rotation rate the vortex
lattice is expected to melt.  The resulting state of matter is
disordered, and it has a large amount of vorticity stored in it.
In this regime, a number of quantum liquid states have been
proposed.  Based on the analogy with the physics of electrons in a
strong perpendicular magnetic field in two dimensions (2D), the
boson quantum liquid states can be characterized as fractional
quantum Hall liquids.

An extensive study of the transition to this regime was conducted
by Cooper, Wilkin, and Gunn \cite{CWG}, who performed exact
diagonalization studies in the LLL in a toroidal geometry. They
predicted that the vortex lattice melts into a quantum disordered
phase at a critical value $\nu_c\simeq 6$--$10$ of the filling
factor $\nu=N/N_v$, with $N_v$ the number of vortices and $N$ the
number of bosons. They also found that the quantum state is
incompressible at $\nu=k/2<\nu_c$, with $k=0$, $1$, $2$, \ldots,
and observed that the ground states at $\nu=k/2$ have substantial
overlaps with the Read-Rezayi (RR) quantum Hall states
\cite{readrezayi1}.

The RR states are incompressible quantum Hall fluids. The special
case $k=1$ is the Laughlin state \cite{laugh} for bosons at
$\nu=1/2$, while the $k=2$ state is the Moore-Read ``Pfaffian''
state \cite{mooreread}. The RR quantum Hall states possess a
specific ``order-$k$'' clustering property: there is a so-called
composite-boson order parameter \cite{gm,read89}, an operator that
creates $k$ bosons and $2$ vortices, and is the minimal order
parameter (the one that creates the smallest number of bosons)
that has long-range (``off-diagonal'') order. Such order implies
that the quasiparticles over these liquids are (quantized)
vortices in the liquid, carrying fractional vorticity $1/k$,
analogous to a fractional magnetic flux $\Phi/\Phi_0$ that occurs
in certain quantum Hall states in electronic systems. In quantum
Hall liquids, the Hall conductivity implies a fundamental
quasiparticle charge $q=\pm\nu\Phi/\Phi_0$ in units of the charge
of the electron.  In the context of neutral bosons in a rotating
trap, the same argument implies that a fractional particle number
($q=\pm1/2$ for the RR states) is present in the quasiparticle,
relative to the background density. Furthermore, for $k>1$ there
are non-local degrees of freedom associated with these
quasiparticles, that is, the ground states with more than three
quasiparticles are degenerate in the limit where all the
separations go to infinity. As these degrees of freedom are of a
nonlocal, topological nature, they do not couple to local probes
and the degeneracy is protected in the large separation limit. For
the case $k=2$ there is an interpretation in terms of a Majorana
fermion in each vortex core \cite{readgreen}\@. Further evidence
for the appearance of the Moore-Read state was recently provided
by Regnault and Jolicoeur \cite{regnault}, who observed the
low-lying two-particle branch in numerical simulations, upon
adding one flux quantum in a spherical geometry. They also found
evidence for other quantum Hall states not in the RR series.

\subsection{Outline}

In previous work \cite{reijnders} we analyzed spin-1 bosons in the
LLL\@.  We identified attractive and repulsive regimes in the
$c_0$--$c_2$ plane, and proposed and analyzed two series of
clustered quantum Hall states (labeled $SU(4)_k$ and $SO(5)_k$),
analogous to the RR states, for spin-1 bosons in a rapidly
rotating trap. We identified the exact ground state for $N$ spin-1
bosons on the disc with one unit of angular momentum per particle,
the boson-triplet-condensate (BTC).

In this paper we provide further results on the phase diagram for
spin-1 bosons in the LLL\@.  Employing the $SU(3)$ symmetry, we
discuss how in a slowly-rotating system with $c_2=0$ the exact
quantum ground state evolves from the non-rotating one towards the
BTC at angular momentum $L=N$, and we compute the ground state
angular momentum $L(\omega)$ as a function of the rotation
frequency $\omega$\@.  Using LLL-mean field theory, we extend the
results for slow rotation to $c_2\neq 0$, and discuss the various
skyrmion lattices.  Furthermore, the two series of quantum Hall
states are discussed in detail and we supplement them with a third
series.

This paper is organized as follows.  In section \ref{II} we define
the model by discussing LLL truncation in a disc, sphere or torus
geometry, specifying the interaction Hamiltonian, and make remarks
on the general symmetry properties for later use. In section
\ref{III}, we study the phase diagram by direct numerical
diagonalization. In section \ref{IV} exact quantum ground state
wavefunctions and energies for a slowly rotating (angular
momentum $L\leq N$) system in the $c_2=0$ limit are presented. For
nonzero $c_2$, we use a LLL mean field treatment to study the
slowly-rotating system (in section \ref{V}) and the various
skyrmion and vortex lattices (in section \ref{VI}).  In section
\ref{VII} we discuss the quantum Hall states at ultra-high
rotation. In an Appendix we discuss the topological classification
of defects.


\section{LLL model Hamiltonian and its symmetry}\label{II}

In this section we describe the truncation of the space of
single-particle states to those in the LLL, and then explain the
use of different geometries (sphere, torus) once this trunction
has been made. Then we give the form of the interaction
Hamiltonian that will be assumed, and some analysis of the
symmetries of the model, with particular reference to certain
limits and different geometries.

\subsection{Truncation to the lowest Landau level}\label{IIA}

In a rotating frame of reference the Hamiltonian for $N$ trapped,
weakly-interacting spin-1 bosons is
\begin{equation}
H=\sum_{i}^N\left[\frac{\omega_0}2(-\vec{\nabla}_i^2+r_i^2)
                  -\vec{\omega}\cdot\vec{L}_i \right] + H_{\rm int}.
\end{equation}
Here $\vec{\omega}$ is the frequency of the rotation drive,
$\vec{L}_i$ the angular momentum of the $i$-th particle and
$H_{\rm int}$ the interaction Hamiltonian, which we discuss below.
We have set $\hbar$ and the harmonic oscillator length
$l\equiv(\hbar/m_b\omega_0)^{1/2}$ of the trap (with $\omega_0$
the trap frequency and $m_b$ the boson mass) equal to one.  Modes
in the direction of the rotation axis are frozen out, leaving us
effectively with a two-dimensional (2D) system.  The energy
eigenvalues of the single-particle part of the Hamiltonian are
then  $E_{n,m}=(2n+m+1) \omega_0-m\omega$, with $n\geq 0$ the
Landau level index and $m\geq -n$ the $z$-component of angular
momentum, labelling the states within each Landau level.

We consider the model in which the single-particle states are
restricted to the lowest ($n=0$) Landau level (LLL) \cite{WG}.
This is valid when the interactions are sufficiently weak, as we
will explain momentarily. The normalized LLL wavefunctions are
$\phi_m(z)\,\zeta^\alpha$ with the orbital part $\phi_m(z)\propto
z^m e^{-|z|^2/2}$ ($z=x+iy$), and $\zeta^\alpha$ a three-component
complex vector representing the spin state; here $\alpha$ labels
the eigenstates of the z-component of the spin $S_z$ for each
particle, $\alpha = \up$, $0$, $\dow$. [Later in the paper it will
be convenient also to use the basis of Cartesian components for
spin 1, labelled by $\mu=x$, $y$, $z$.] When we use second
quantization, we will denote the boson creation and annihilation
operators for these single-particle states by
$b_{m\alpha}^\dagger$, $b_{m\alpha}$, and the corresponding
occupation numbers by $n_{m\alpha}\equiv b_{m\alpha}^\dagger
b_{m\alpha}$. Also, we sometimes use the field operator
$\psi_\alpha(z)=\sum_m b_{m\alpha}\phi_m(z)$\@. The single
particle contributions to the Hamiltonian add up to
$(\omega_0-\omega)L$, with $L=\sum_i L_{zi}$ the $z$-component of
total angular momentum.  We will refer to this geometry as the
disc in view of the form of the fluid states (for repulsive
interactions) which tend to form a disc or ``pancake'', because of
the centrifugal force. Note that we must have $\omega\leq
\omega_0$, otherwise the system becomes unstable.

To study the bulk properties of the quantum ground states, we will
eliminate boundary effects by using instead two other geometries
and taking the limit $\omega\rightarrow\omega_0$\@.  In a
spherical geometry \cite{haldane}, the orbital part of the LLL
single-particle wavefunctions is $\phi_m(z) \propto
z^m/(1+(\frac{|z|}{2R})^2)^{1+N_v/2}$, where $z$ represents
position on the sphere by stereographic projection to the plane,
and $R$ is the radius of the sphere. The number of orbitals is
restricted by the vorticity $N_v$ penetrating the sphere, $0\leq
m\leq N_v$\@. The $N_v+1$ single-particle orbitals form a
representation of orbital angular momentum equal to $N_v/2$, see
Ref.\cite{haldane}.  In the limit $R\rightarrow\infty$, keeping
$N_v/R^2$, $N$ and $z$ constant, the single-particle wave
functions on the sphere reduce to those for the disc as above. The
total angular momentum on the sphere is characterized by quantum
numbers $\tilde{L}$ for the magnitude, and $\tilde{L}_z$ for the
$z$-component. In terms of $L$ which has eigenvalues $L=\sum_i
m_i$ as before, $\tilde{L}_z={1 \over 2} NN_v-L$\@. We emphasize
that our definition of $L$ when used for the sphere does not have
its usual meaning, but is related to the $z$-component in such a
way that the $N_v\to\infty$ limit agrees with the plane.

The final geometry we use is the torus. Here the single-particle
wavefunctions take the form $\phi(z)\propto f(z)e^{-y^2}$ in the
Landau gauge, with $f$ a quasiperiodic holomorphic function. With
$N_v$ flux quanta, $f$ has $N_v$ zeros in the unit cell. There are
exactly $N_v$ independent solutions, of the form
$f(z)=\prod_{i=1}^{N_v} \vartheta_1(z-z_i|\tau)$, with $\tau$
describing the geometry of the unit cell and $z_i$ the zeros of
$f$\@. The use of $\vartheta$-functions ensures that $\phi$ is
periodic. Many-body states can be classified by their Haldane
momentum \cite{haldane2}.

\subsection{Interaction Hamiltonian}

In a model description, the Hamiltonian describing the 2-body interactions
of a system of $N$ spin-1 bosons is a contact interaction, and contains
spin-independent ($H_n$) and spin-dependent ($H_s$) terms, of strengths
$c_0$, $c_2$ respectively
\begin{eqnarray}
H_{\rm int}&=& H_n+H_s
\nonumber \\
&=& 2\pi\sum_{i<j}^N\delta^{(2)}({\bf r}_i - {\bf r}_j)
    \left[ c_0 + c_2 \, \vec S_i\cdot\vec S_j \right] .
\label{hamiltonian}
\end{eqnarray}
Here $c_0=(g_0+2g_2)/3$, $c_2=(g_2-g_0)/3$,
$g_S=4\pi\hbar^2a_S/m_b$ and $a_S$ ($S=0$, $2$) the 2D $s$-wave
scattering phase shift in the spin-$S$ channel \cite{OM,Ho}\@. A
factor $2\pi$ has been extracted for later convenience.  One can
obtain these parameters by integrating over the third direction.
Assuming, for example, harmonic confinement with quantum length
$l_\perp$ in the $z$-direction, one finds
$a^{2D}_S=a^{3D}_S/\sqrt{2\pi}l_\perp$ when $l_\perp\ll l$. For
the sphere, the coordinates $\bf r$ in this Hamiltonian take
values on the surface of the sphere, with radius $R$.

The use of the LLL reduced Hamiltonian is justified when the
interactions are weak. Physical quantities evaluated in the full
model differ from those in the LLL model by relatively small
corrections when the $\nu c_S \ll 2\omega_0$. Here $\nu$ is the
typical filling factor (expectation of the occupation numbers,
summed over $\alpha$ or $\mu$) of the single-particle states.
Notice that this condition becomes much less stringent as
$\omega\to\omega_0$ in the repulsive regime, as then the particles
spread out into a pancake, and the filling factor $\nu$ becomes of
order $1$.

Finally then, the LLL Hamiltonian in the rotating frame which we
wish to analyze is
\begin{equation}
H_{\omega} = (\omega_0 - \omega)L + H_{\rm int} \ .
\label{H-omega}
\end{equation}
Note that we use precisely this definition in the case of the
sphere as well as for the disc. It will be useful also to know the
ground states of $H_{\rm int}$ for each $L$.

\subsection{$SU(3)$ symmetry analysis for $c_2=0$}

In general, the only symmetry in spin space of the Hamiltonians
$H_{\rm int}$ and $H_\omega$ is spin-rotation symmetry $SO(3)_{\rm
spin}$. This implies that spin states will come in multiplets of
spin $S$ with degeneracy $2S+1$ (with $S$ integer since the
particles have spin 1). However, at $c_2=0$, the interaction
Hamiltonian reduces to the spin-independent interaction $H_n$\@.
In this case the spin-rotation symmetry is enlarged from
$SO(3)_{\rm spin}$ to $SU(3)_{\rm spin}$. It will be useful to
understand what this implies about the spin multiplets in a finite
size system.

For $c_2=0$, the spectrum will contain degenerate spin multiplets
labelled by $SU(3)$-quantum numbers $(p,q)$\@.  These tuples are
the Dynkin indices labelling irreducible representations of
dimension ${\rm dim}_{(p,q)}=\frac12(p+1) (q+1)(p+q+2)$\@. Since
$SO(3)$ is embedded in $SU(3)$, each multiplet can be decomposed
into a set of $SO(3)$ multiplets.  These $SO(3)$ spin quantum
numbers can be deduced by using branching rules for $SU(3)\mapsto
SO(3)$\@.  The fundamental branching rule states that a $(p,0)$ or
$(0,p)$ multiplet contains $S = p$, $p-2$, $p-4$, \ldots, $1$
($0$) for $p$ odd (resp., even). Using the fusion rule
\begin{eqnarray}
\lefteqn{(p,0)\otimes(0,q)} \nonumber \\
&& =\,(p,q)\oplus(p-1,q-1)\oplus\cdots\oplus(p-q,0) \ ,
\end{eqnarray}
which is valid for $p\geq q$, general branching rules can be
derived. A multiplet $(p,q)$ with $q$ odd and $p\geq q$ decomposes
in $SO(3)$ multiplets with highest weights $S$ according to the
branching rule
\begin{equation}
(p,q)\,\mapsto\,
\bigoplus_{i=0}^{\frac{q-1}{2}}\bigoplus_{S=2i+1}^{p+q-2i}S\quad.
\end{equation}
For $q$ even we find
\begin{eqnarray}
(p,q)&\mapsto&
\bigg(\bigoplus_{i=0}^{\frac{q-2}{2}}\bigoplus_{S=2i+1}^{p+q-2i}S\bigg)
    \oplus\bigg(\bigoplus_{j=\frac{q+1}{2}}^{\frac{p}{2}}2j\bigg),
    \quad p {\rm ~odd}\quad\ \\
(p,q)&\mapsto&
\bigg(\bigoplus_{i=0}^{\frac{q-2}{2}}\bigoplus_{S=2i+2}^{p+q-2i}S\bigg)
    \oplus\bigg(\bigoplus_{j=0}^{\frac{p}{2}}2j\bigg),
    \qquad p {\rm ~even}.
\end{eqnarray}
Note that the highest $SO(3)$-spin in an $SU(3)$-multiplet $(p,q)$
is always $S=p+q$, and the lowest $S=0$ or $1$.

\subsection{Orbital symmetry in spherical geometry}

In the plane geometry, $H_{\rm int}$ is invariant under
translations and rotations in the plane. When working on the
sphere, this symmetry group is replaced by the rotation group
$SO(3)_{\rm orb}$ (strictly, we should say $SU(2)_{\rm orb}$
whenever $N_v$ is odd) of the sphere. In the limit
$R\rightarrow\infty$ described above, this symmetry becomes
translations and rotations of the plane. When taking this limit,
we also hold $L$ fixed, and hence many-particle states of definite
$SO(3)_{\rm orb}$ quantum numbers $(\tilde{L},\tilde{L}_z)$ become
in the limit infinite-dimensional multiplets of the Euclidean
group of the plane. States within each multiplet differ only in
the state of the center of mass variable (which has coordinate
$z_c=\sum_i z_i/N$). Thus, if $\psi_L$ is an eigenfunction of
$H_{\rm int}$ at certain angular momentum $L$, then there exists a
whole ``tower'' of states $\psi_{L+L'}\propto z_c^{L'}\psi_L$ with
the same interaction energy at angular momentum $L+L'$\@.

We remark that in situations where only a few quantum orbitals are
available to the bosons, the spectrum is largely determined by
symmetry considerations. Particular examples are the spectrum for
$N_v=2$ on the sphere, where the exact $N$-body energies are given
in terms of Casimir invariants of the orbital and spin-symmetries
(see eq.~(\ref{en_sphere}) below), and the case with $N_v=4$ on
the torus, where the topological degeneracy eq.~(\ref{torus-deg})
below pertaining to particular quantum liquid states is recovered
from the $SU(3)$ spin-symmetry.


\section{Main features of the phase diagram}\label{III}

In this section we make a first pass through the phase diagram
with numerical results on moderate sizes. First we consider the
ground states of $H_{\rm int}$ in the disc geometry for each $L$,
then use this to find the ground states of $H_\omega$ as a
function of $\omega$. All this has to be done for general values
of $c_0$, $c_2$.

\begin{figure}
\epsfig{file=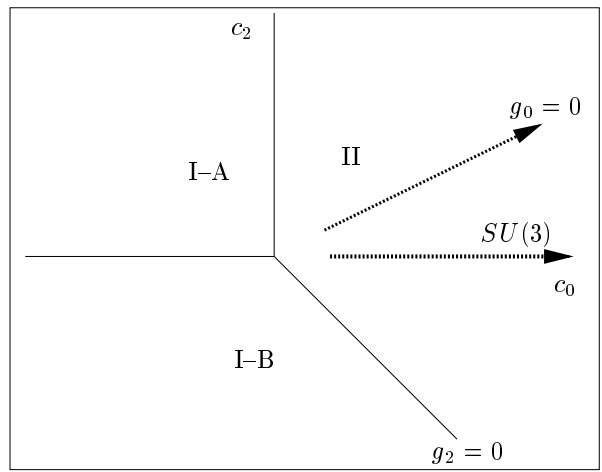,width=85mm} \caption{Overview of
$c_0$--$c_2$ plane, with special regions and directions marked.}
\label{fig:c-plane}
\end{figure}

\subsection{Global structure of the phase diagram}

First we point out that the magnitude $(c_0^2+c_2^2)^{1/2}$ only
sets the overall energy scale, so it can be divided out. Thus the
phase diagram can be thought of as a circle, in which a point on
the circle represents a ray in the $c_0$--$c_2$ plane. We wish to
examine this for each $L$, or later for each $\omega$. In
Figure~\ref{fig:c-plane} the $c_0$--$c_2$ plane is shown with
certain special directions ($c_0=0$, $c_2=0$, $g_0=0$, $g_2=0$)
that will be important later picked out.

For $L=0$, the ground state has total spin $S=N$ for $c_2<0$
(ferro regime) and $S=0$ ($1$) for $N$ even (odd) for $c_2>0$
(anti-ferro regime)\@. These states are the way that the broken
symmetry states described in Sec.~\ref{I} (the ferromagnetic and
polar states respectively) appear in a finite size study. For
$c_2=0$, there is a single $SU(3)$ multiplet of spin states,
decomposing into one $SO(3)$ multiplet of each spin $S=N$, $N-2$,
\ldots\phantom{.} The transition at $c_2=0$ can thus be viewed as
levels crossing, with a larger degeneracy on the line $c_2=0$. As
$L$ increases, these two phases at $c_2\neq0$ survive in part of
the phase diagram, as compact drops of fluid, with the center of
mass carrying all the angular momentum. Meanwhile, the positive
$c_0$ axis gradually opens into a region that contains other
phases.  By the time $L$ is $\geq N$, the $c_0$--$c_2$ plane
contains the three regions labelled I--A, I--B, and II in
figure~\ref{fig:c-plane}\@.

The ground states in regions I--A and I--B are similar to the one
in the ``attractive'' regime in the scalar case \cite{WGS}\@.  The
orbital part of the ground state wavefunction is of the form
$\tilde{\Psi}(z_i) \propto z_c^L$. In region I--A ($c_0<0$,
$c_2>0$), the spin state is the same spin-singlet as for the $L=0$
ground state, and the energy \cite{ho-yip} becomes $[c_0 N(N-1)/2
- N c_2]$\@. In region I--B ($c_2<0$, $c_0<-c_2$), the spin state
is ferro, $S=N$, giving energy \cite{WGS} $(c_0+c_2)N(N-1)/2 <
0$\@.  At $c_2=0$, $c_0<0$, the spin states again form the $SU(3)$
multiplet.  In the remaining ``repulsive'' region II, the ground
state is in general not a common eigenstate of the $c_0$ and $c_2$
parts of the interaction, and the ground state energy depends
non-linearly on the ratio $\gamma=c_2/c_0$. Note that we have now
located the repulsive region more precisely than in our previous
characterization of it simply as $c_0>0$. Most of the following
analysis focuses on region II only, which can be parametrized by
$\gamma=c_2/c_0$ alone.

\subsection{Finite size results in region II as a function of $\omega$}

\begin{figure}[!ht]
\epsfig{file=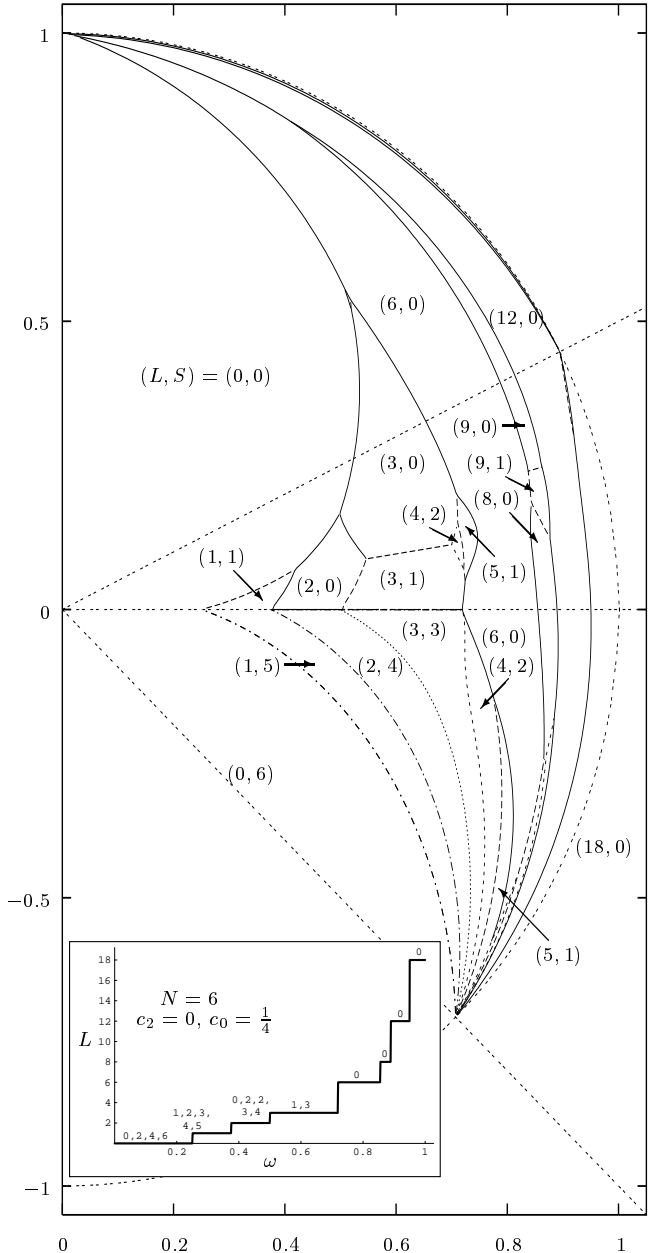,width=88mm}
\begin{center}
\hskip2mm \caption{Ground state quantum numbers $(L,S)$ in region
II for $N=6$ spin-1 particles in the planar (disc) geometry, as a
function of the driving frequency $\omega$ (plotted radially) and
the ratio $\gamma$ (corresponding to the angle with respect to the
horizontal axis). The special directions $g_2=0$, $c_2=0$,
$g_0=0$, $c_0=0$ are shown as double-dotted radial lines. The
inset shows a cut along the $c_2=0$ direction, with the angular
momentum given on the vertical axis and the (degenerate) spin
values $S$ marked at each of the steps. In this figure, the
parameters $c_0$, $c_2$, and $\omega$ are in units of $\omega_0$,
and the value $c_0=0.25$ is used in the main figure as well as in
the inset. For additional discussion, see the main text.}
\label{pl}
\end{center}
\end{figure}

In figure~\ref{pl} we show the ground state quantum numbers
$(L,S)$ in region II for $N=6$ bosons as a function of the
rotation frequency $\omega$. As the phase diagram for each
$\omega$ is a circle (which in region II can be parametrized by
$\gamma=c_2/c_0$, or by $\phi=\arctan \gamma$), we are free to
plot $\omega$ radially. The parameters are shown in units of
$\omega_0$ and with $c_0=0.25$, but notice that the ground state
quantum numbers can only depend on the dimensionless ratios of
energies $(\omega_0-\omega)/c_0$ and $c_2/c_0$, so that the
structure shown is actually present (though with the radial
variable rescaled and shifted) for all parameter values (unless
$c_0$ is too large). The dashed rays are the lines $c_2=-c_0$,
$c_2=0$ and $c_2=c_0/2$, and the outer dashed circle is the locus
of $\omega=\omega_0$. The ground state angular momentum $L$ and
spins $S$ at $c_2=0$ and as a function of $\omega$, are shown in
the inset. The degenerate spin values at $L \leq N$ are seen to
correspond to the following irreducible $SU(3)$ multiplets:
$(p,q)=(6,0)$ for $L=0$, $(p,q)=(4,1)$ for $L=1$, $(p,q)=(2,2)$
for $L=2$, $(p,q)=(1,1)$ for $L=3$ and $(p,q)=(0,0)$ for $L=6$.
Note also that for $c_2<0$, $c_0>-c_2$ the ground state spin
gradually decreases from $S=6$ at $\omega=0$ to $S=0$.


\section{Slow rotation: exact ground states at $c_2=0$, $L\leq N$}
\label{IV}

For larger sizes, a brute-force numerical approach is not
feasible, so we develop other approaches. In this section we
determine the exact ground state energies and wavefunctions for
slow rotation (angular momentum up to the boson number, $L\leq
N$), for $c_0>0$ and $c_2=0$, exploiting the $SU(3)$ symmetry
described in the section~\ref{II}. Some of the ground states we
find were described in Ref.\ \cite{hocluster}. We analyze a system
of $N$ spin-1 bosons in spherical geometry with $N_v$ quanta of
vorticity, with the disc geometry emerging as the limit
$N_v\rightarrow\infty$\@. We remark that for $N$ sufficiently
large, it becomes natural to discuss low energy properties in
terms of mean field configurations that break the various
symmetries and whose energy is slightly higher than that of the
exact quantum ground state; this will be discussed in section
\ref{V}.

\subsection{Exact eigenstates of $H_n$}

The ground state spectrum for $c_2=0$ and $L\leq N$ can be
understood by exploiting the $SU(3)$ symmetry of the Hamiltonian
$H_n$\@.  In our analysis we proceed as follows.  We consider two
series of eigenstates of $H_n$, in which (roughly speaking) the
bosons occupy at most the lowest three orbitals. Among these
eigenstates, we identify the exact quantum ground states on the
disc and the sphere, as a function of the angular momentum. This
then allows us to compute the $\omega$ dependence of the ground
state angular momentum for general $N$ at $c_2=0$.

We write the first series of eigenstates as $|p,q,n\ket^I$\@.
These states contain doublets and triplets of spin-1 bosons that
are fully antisymmetric in spin indices, and in the orbital
indices (guaranteeing the overall symmetry that is required). The
different numbers of single bosons, doublets and triplets
correspond uniquely to the values of $N$ and the quantum
numbers $(p,q)$ of the corresponding $SU(3)$ multiplets.  The
triplets, which appear $n$ times, are singlets under $SU(3)$, and
so do not affect the overall $SU(3)$ representation. The highest
spin component ($S^z=p+q$) of the corresponding $SU(3)$ multiplet
takes the following form (up to normalization)
\begin{eqnarray}
\lefteqn{|p,q,n\ket^I\,\propto}\nonumber\\
&&[\vec{e}_1\cdot \vec{B}_\up^\dagger]^p[\vec{e}_2
\cdot(\vec{B}^\dagger_\up\times\vec{B_0^\dagger})]^q[\vec{B}_0^\dagger
\cdot(\vec{B}_\up^\dagger\times\vec{B}_\dow^\dagger)]^n|0\ket,\quad
\label{wavef}
\end{eqnarray}
with $\vec{e}_1=(1,0,0)$, $\vec{e}_2=(0,0,1)$ and
$\vec{B}^\dagger_\alpha= (b^\dagger_{0,\alpha},
b^\dagger_{1,\alpha}, b^\dagger_{2,\alpha})$\@.  Clearly, the
total number of bosons is $N=p+2q+3n$.  The energies corresponding
to eq.~(\ref{wavef}) are
\begin{eqnarray}
E_{p,q,n}^I/c_0&=&\alpha_1^{N_v}n(n-1)+\alpha_2^{N_v}q(q-1)+
\frac{1}{2}p(p-1)\nonumber\\
&&+\alpha_3^{N_v}np+\frac{3}{2}qp+\alpha_4^{N_v}nq
\label{Nvenergy},
\end{eqnarray}
with
\begin{eqnarray}
&\alpha_1^{N_v}=3\frac{11N_v^2-20N_v+6}{4(2N_v-3)(2N_v-1)}\quad
\alpha_2^{N_v}=\frac{5N_v-2}{2(2N_v-1)}&\nonumber\\
&\alpha_3^{N_v}=\frac{7N_v-4}{2(2N_v-1)}\qquad\alpha_4^{N_v}=
\frac{5\alpha_2^{N_v}}{2}.&\nonumber
\end{eqnarray}
This energy is for spherical geometry, and it depends on the number
$N_v$ of flux quanta.  For $N_v\rightarrow\infty$ eq.~(\ref{Nvenergy})
gives the energy in a disc geometry; $N_v=2$ gives the energy on a
sphere with 3 orbitals.  On the basis of exact diagonalization studies
for $N=6,9,12,15,18$ particles we claim that on the disc for $L\leq
N/2$, the ground state multiplet is precisely $|p,q,0\ket^I$, with
$p=N-2L$, $q=L$.

On the sphere with $N_v=2$, we have obtained a much stronger
result \cite{thanksEd}, namely a closed form result for {\it all}
eigenvalues of $H_n$\@.  It turns out that these energies can be
given in terms of the number $N$ of bosons, the total angular
momentum $\tilde{L}$ and the $(p,q)$ labels of the $SU(3)$
representation, according to
\begin{eqnarray}
E_{p,q}^{N_v=2}/c_0&=&\frac{5}{18}N(N-1)+\frac{1}{6}T^2_{p,q}+\frac{1}{6}
\tilde{L}(\tilde{L}+1),
\nonumber\\ &&
\label{en_sphere}
\end{eqnarray}
where $T^2_{p,q}=(p^2+q^2+pq)/3+p+q$ is the quadratic Casimir operator
for $SU(3)$ in the representation $(p,q)$\@.  Specializing this
expression to the states in series I, by eliminating $N$ in favor of $n$
and using the fact that $\tilde{L}=p+q$, reproduces the result in
eq.~(\ref{Nvenergy}) for $N_v=2$.

Analyzing the ground state on the disc for $L>N/2$, we identified
a second series of states $|p,q,n\ket^{\it II}$. One can think of
the type II states as having the $p$ single bosons in $m=1$ rather
then $m=0$, so that now $\vec{e}_1=(0,1,0)$. That is not quite
correct for the energy eigenstates, as we will explain below, but
it does give the correct quantum numbers.  The states in series I,
II share the property of having $p$ single bosons and $q$
doublets, leading to $SU(3)$ Dynkin labels $(p,q)$\@.  It may be
illuminating to display the structure of the states in terms of
diagrams similar to Young tableaux as in figure \ref{plaatje}. For
the orbital structure of the highest-weight states in either
series I or II, the lengths of the three rows represent the number
of bosons in the orbitals $m=0$, $1$, $2$ respectively (in the
rough point of view, which will be corrected below), while the
differences $p$, $q$ and $n$ in the lengths correspond to the
$SU(3)$ structure. Essentially, these diagrams are ordinary Young
tableaux for the states, but with the first two rows exchanged in
the case of series II.

\setlength{\unitlength}{1.5em}
\def\boxje{\framebox(.473,.473){}}
\begin{figure}
\begin{picture}(15,4)(3.5,4)
\put(4.75,5.25){$n$}
\put(5,5.75){\vector(1,0){1}}
\put(5,5.75){\vector(-1,0){1}}
\put(7,6.25){\vector(1,0){1}}
\put(7,6.25){\vector(-1,0){1}}
\put(6.75,5.7){$q$}

\put(5,4.2){$|p,q,n\ket^{\it I}$}
\multiput(4,7)(.5,0){2}{\boxje}
\multiput(5.5,7)(.5,0){3}{\boxje}
\multiput(7.5,7)(.5,0){3}{\boxje}
\multiput(9.5,7)(.5,0){1}{\boxje}

\put(5,6.5){..}
\put(7,7){..}
\put(9,7.25){..}

\put(9,6.75){\vector(1,0){1}}
\put(9,6.75){\vector(-1,0){1}}
\put(8.75,6.2){$p$}

\multiput(4,6.5)(.5,0){2}{\boxje}
\multiput(5.5,6.5)(.5,0){1}{\boxje}
\multiput(6,6.5)(.5,0){2}{\boxje}
\multiput(7.5,6.5)(.5,0){1}{\boxje}

\multiput(4,6)(.5,0){2}{\boxje}
\multiput(5.5,6)(.5,0){1}{\boxje}


\put(12.75,5.25){$n$}
\put(13,5.75){\vector(1,0){1}}
\put(13,5.75){\vector(-1,0){1}}
\put(15,6.25){\vector(1,0){1}}
\put(15,6.25){\vector(-1,0){1}}
\put(14.75,5.7){$q$}

\put(13,4.2){$|p,q,n\ket^{\it II}$}
\multiput(12,7)(.5,0){2}{\boxje}
\multiput(13.5,7)(.5,0){1}{\boxje}
\multiput(14,7)(.5,0){2}{\boxje}
\multiput(15.5,7)(.5,0){1}{\boxje}
\multiput(16,6.5)(.5,0){2}{\boxje}
\multiput(17.5,6.5)(.5,0){1}{\boxje}

\put(13,6.5){..}
\put(15,7){..}
\put(17,6.75){..}

\put(17,6.25){\vector(1,0){1}}
\put(17,6.25){\vector(-1,0){1}}
\put(16.75,5.7){$p$}

\multiput(12,6.5)(.5,0){2}{\boxje}
\multiput(13.5,6.5)(.5,0){1}{\boxje}
\multiput(14,6.5)(.5,0){2}{\boxje}
\multiput(15.5,6.5)(.5,0){1}{\boxje}

\multiput(12,6)(.5,0){2}{\boxje}
\multiput(13.5,6)(.5,0){1}{\boxje}

\end{picture}

\caption{The structure of the two different series of eigenstates
of $H_n$, displayed in a form similar to Young tableaux. In both
cases, the corresponding $SU(3)$-representation has Dynkin labels
$(p,q)$.} \label{plaatje}
\end{figure}
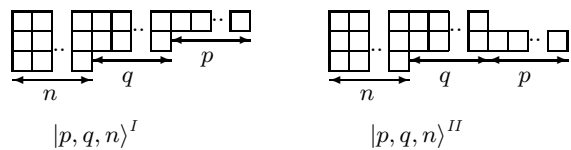

For the case of the type II states, the following correction must
be made to obtain the energy eigenstates. In the case of scalar
bosons, it is known\cite{WG,BP,wil-sm} that the ground state
configuration at $L=p$ of $p$ bosons is a vortex located at their
center of mass, with wavefunction $\prod_i (z_i-z_c)$ with $z_c =
\sum_i z_i/p$\@.  This state is not entirely restricted to the
$m=1$ orbital, as there are components in which other orbitals in
the range $0\leq m \leq p$ are occupied as well. The $p$ bosons in
the state $|p,q,n\ket^{\it II}$ form such a vortex. This
complication makes it difficult to write down the closed form
expression for the states in series II; based on numerical
analysis for small $N$ and mean field results for large $N$ (see
section \ref{V}), we do propose the following closed form
expression for the corresponding energy on the disc
\begin{eqnarray}
E_{p,q,n}^{\it II}/c_0&=&\frac{33}{16}n(n-1)+\frac{5}{4}q(q-1)+
\frac{1}{4}p(p-2)\nonumber\\
&+&\frac{25}{8}nq+\frac{11}{8}np+qp.
\label{energy2}
\end{eqnarray}
Note that the $p$-independent terms in this formula are identical
to those for type I states with $N_v=\infty$.  The state
$|p,0,0\ket^{\it II}$, has energy $p(p-2)/4$, which is exactly the
ground state energy of a rotating scalar BEC at $L=p=N$. This
justifies the interpretation of the polarized subsystem with $p$
bosons forming a vortex at the center of mass\@.  However, it
turns out that $|p,0,0\ket^{\it II}$ will never be the lowest
energy configuration for a rotating spin-1 system.

Among the type I/II states the following are special. First,
$|p,0,0\ket^{\it I}$ is the non-rotating ground state,
corresponding to the $(p,0)$-multiplet.  Second, $|0,q,0\ket$
gives a wavefunction composed of anti-symmetrized pairs of
bosons, a Boson-Doublet-Condensate (BDC) or $(0,q)$-multiplet.
Third, $|0,0,n\ket$ is composed of 3-body singlets. It is a
condensate of triplets or boson-triplet-condensate
\cite{reijnders} (BTC); we shall see that it forms the ground
state at $L=3n=N$\@. The BTC-state can be regarded as a
symmetrized version of the core-less vortices observed in mean
field studies (see section \ref{V} for more on this).

More generally, the type I/II states are examples of ``(multi-)
fragmented'' condensates \cite{noz-james}, see also \cite{ho-yip},
in the sense that they contain several macroscopically occupied
elements in the density matrix.  For instance, for the BTC- and
for (any component of) the BDC-state we have $\bra
n_{m\alpha}\ket_{\rm BDC}=(1-\delta_{m,2})
(1-\delta_{\alpha\dow})q/2$, $\bra n_{m\alpha}\ket_{\rm
BTC}=n/3$\@.  Since the spin is fixed in these states, $(\Delta
n_\alpha)^2\equiv \bra (n_\alpha-\bra n_\alpha\ket)^2\ket=0$,
where $n_\alpha=\sum_{m=0,1,2}n_{m\alpha}$. However, within each
spin component, the fluctuations of the boson number between
orbitals is of the order of the system size: $(\Delta
n_{m\alpha})_{\rm BDC}^2=q(q+2)/12$, $(\Delta n_{m\alpha})_{\rm
BTC}^2=n(n+3)/18$. This is an indication that, as in the case of
the singlet ground state at $L=0$ in the antiferromagnetic regime
and the related ``polar'' mean field state \cite{ho-yip}, it may
be best to think of these states as broken symmetry states
\cite{And}. That is the approach we will take in Sec.~\ref{V}.

\begin{figure}
\begin{center}
\epsfig{file=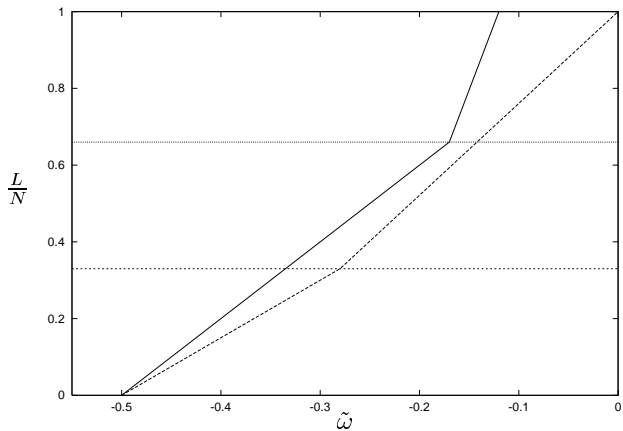,width=8.5cm} \caption{$L/N$
as $N\to\infty$ of the ground state on the disc (bold line) and on
the sphere with $N_v=2$ (dashed line), as functions of
$\tilde{\omega}= (\omega-\omega_0)/(c_0N)$ at $c_2=0$. The
horizontal lines mark the values $L/N=1/3$ and $L/N=2/3$\@. The
cusps in both curves indicate the point where the $m=2$ quantum
orbital is first used in the ground state.} \label{plaatje1}
\end{center}
\end{figure}

\subsection{Exact ground states at $c_2=0$ as a function of $L$ or $\omega$}

The ground state of a rotating gas with $N$ spin-1 bosons in the
LLL and a spin-independent ($c_2=0$)  interaction is formed by a
sequence of type I or II states lying on  a certain path in
$(p,q)$-space as $L$ increases. To find the ground state in a
rotating frame of reference, we need to find the ground state of
$H_\omega$, eq.\ (\ref{H-omega}), instead. Since this Hamiltonian
contains only two energy scales, the ground state angular momentum
per particle $L/N$ can be written as a function of the ratio
$(\omega-\omega_0)/(Nc_0)$\@. For finite boson number this
function consists of a sequence of steps, as can be seen in figure
\ref{pl}. It turns out that (thanks to our judicious choices of
factors of $N$) the limit $N\to\infty$ with $L/N$ and
$(\omega-\omega_0)/(c_0N)$ fixed of this function exists, and this
is the most convenient information to display. In the following we
determine the path of the ground states in $(p,q)$-space as a
function of $L$, and the $L(\omega)/N$ behavior of the ground
states in this limit for both the sphere ($N_v=2$) and the disc
($N_v=\infty$) in the regime $L/N\leq 1$.

\subsubsection{Ground states on the sphere at $N_v=2$}

On the sphere, our notion of rotation is such that the $SO(3)_{\rm
orb}$-angular momentum $\tilde{L}$ decreases as the system rotates
faster and faster. With three orbitals ($N_v=2$) we have
$\tilde{L}_z=N-L$\@ (see section \ref{IIA}). (We consider $N_v=2$
because this case can just accommodate $L\leq N$.) At
$\tilde{L}=N$, we know already that the $|N,0,0\ket^{\it I}$
multiplet forms the ground state\@. As $\tilde{L}$ starts to
decrease, again a type I state has the lowest energy; the
$(p,q)$-path is parametrized by $(2\tilde{L}-N,N-\tilde{L})$\@.
Bosons are gradually added to the $m=1$ orbital and form
anti-symmetrized pairs with the remaining ones. The point up to
which this continues can be found by comparing the energies of
$|2\tilde{L}-N,N-\tilde{L},0\ket^{\it I}$ and
$|2\tilde{L}-N+t,N-\tilde{L}-t,t/3\ket^{\it I}$\@.  After
minimizing with respect to $t$ this yields the critical
$SU(3)$-indices $(p,q)_c=(N/3,N/3)$\@. At this point, with
$\tilde{L}=2N/3$, ground states with a nonzero $(n>0)$ number of
triplets become energetically favorable.  In the remaining region,
$2N/3\geq\tilde{L}\geq 0$, type I states are the ground states
following the path $(p,q)=(\tilde{L}/2,\tilde{L}/2)$\@. Eventually
this terminates on the BTC at $\tilde{L}=0$\@. $L/N$ of the ground
state as a function of the rotation drive $\omega$ shows a cusp at
$\tilde{L}/N=2/3$ ($L/N=1/3$), as is shown in
figure~\ref{plaatje1}.

\subsubsection{Ground states on the disc}

For a system on the disc ($N_v=\infty$), the results are rather
different. We will again present the ground states in order of
increasing $L$. At  $L=0$,  the $|N,0,0\ket^{\it I}$-multiplet
forms the ground state as we know.  For $L\leq N/2$ the ground
state is formed by a type I state with $n=0$ and $SU(3)$-quantum
numbers $(p,q)=(N-2L,L)$. This state terminates on the BDC at
$L=N/2$\@. In this range, increasing $L$ leads, as on the sphere,
to more bosons occupying the $m=1$ orbital, forming
anti-symmetrized pairs with the ones in the $m=0$ orbital. For
$L\geq N/2$, the type II states have the lowest energy.  As $L$
increases, bosons move from the $m=0$ into the $m=1$ orbital,
decreasing the number of doublets, and giving type II states at
$(p,q)=(2L-N,N-L)$\@. Comparing the energies of
$|tN,(1-t)N/2,0\ket^{\it II}$ and $|(t-s)N,(1-t)N/2, s
N/3\ket^{\it II}$, we can determine the point where it becomes
favourable for triplets to enter the ground state.  We find a
critical angular momentum $L=(1-t_c)N$ with $t_c\sim 1/3-3/N$,
which approaches $L=2N/3$ for $N$ large.  For $L\geq 2N/3$ the
number of triplets is gradually increasing as $L$ grows.
Minimizing $|2L-N-s,N-L,s/3\ket^{\it II}$ with respect to $s$, we
find that the ground state is now the type II state with
$s(N,L)=3L-2N$, giving $(p,q,n)=(N-L,N-L,L-2N/3)$\@.  For $L=N$
the ground state is the BTC with $p=q=0$, $n=N/3$.

To summarize the above results, for $0\leq L\leq N/2$ the ground
state is given by $|N-2L,L,0\ket^{\it I}$ and for $N/2\leq L\leq
2N/3$ by $|2L-N,N-L,0\ket^{ \it II}$\@.  In the remaining range
$2N/3\leq L\leq N$ the number of 3-body singlets is nonzero, and
the ground state is given by $|N-L,N-L,L-2N/3\ket^{\it II}$\@.
Minimizing the energy in a rotating frame of reference leads to
the $L(\omega)/N$-dependence of the ground states for $N\to\infty$
which is depicted in figure \ref{plaatje1}. In this figure, the
curve shows a cusp at the point where the $m=2$ orbital first
enters the ground state configuration, which is at $L/N=2/3$ for
the disc with $N$ large. A signature of this cusp in an
experimental system might be a change in the expansion rate (the
rate of change of the outer radius of the drop with respect to
$\omega$) if the angular momentum exceeds $2N/3$. We shall see
that the cusp survives in the anti-ferromagnetic regime, $c_2>0$.

It is important to contrast all this with the well-known behavior
of scalar bosons in a rotating trap \cite{WG,WGS}. In the latter
case there is a jump from $L/N=0$ to $L/N=1$ (for all $N$) when
one vortex enters the system, whereas for spin-1 bosons we find (at 
$N\to\infty$) a continuous $L(\omega)/N$ curve with a discontinuous 
slope.


\section{Slow rotation: LLL mean field theory}
\label{V}

At low rotation rates, the typical boson occupation numbers
$\langle n_{m\alpha}\rangle$ of the occupied ($n_{m\alpha}\neq 0$)
single-particle states are large compared with 1. In this
situation, a mean field (or classical) approach to the problem is
generally expected to be quantitatively accurate. In such an
approach, the boson operators are replaced by expectation values,
which are complex c-numbers: $b^\dagger_{m\alpha}\rightarrow
b_{m\alpha}^*$, and the second-quantized Hamiltonian is then
minimized with respect to both the magnitude and phase of these
numbers to find the ground states. In essence the resulting state
is a Bose condensate with the bosons condensed in one linear
combination of the single-particle states. This typically involves
breaking the orbital and spin symmetries, as well as particle
number conservation. (States with definite values of the good
quantum numbers such as $N$, $S$, $L$ can be obtained afterwards
by applying a projection to the mean field quantum state \cite{And}.) 
In the case of very low rotation, where $L\leq N$, we have seen that
(neglecting the subtleties that arose for type II states) the
states essentially involve only the $m=0$, $1$, and $2$ states, so
that the basis set for the mean field calculation is particularly
small. In these cases, the mean occupation numbers of the single
particle states are of order $N$, and their energies exceed the
exact ground state energy (which is of order $N^2$) by an amount
of order $N$. In this section we pursue this mean field
calculation for this regime. This gives us easy access to the
ground states at large $N$ for $c_2\neq 0$ in region II\@. In the
following section, we study instead the mean field states at larger
rotation, which can be assumed to be states in which the translational
and rotational symmetry group of the plane is broken to that of a
lattice.

In terms of the complex numbers $b_{m\alpha}$, $b_{m\alpha}^\ast$,
the energy becomes a quartic polynomial and the ground state can
be found by minimizing this polynomial with respect to these
variables. This is done here with the {\em mean} boson number
$\langle N\rangle=\sum_{m\alpha} b_{m\alpha}^\ast b_{m\alpha}$ and
angular momentum $\langle L\rangle=\sum_{m\alpha}m
b_{m\alpha}^\ast b_{m\alpha}$ fixed at the values $N$, $L$,
respectively. The spin is not constrained at all. For $c_2=0$ the
Hamiltonian on the sphere takes the form
\begin{equation}
H_n=c_0\sum_{\alpha\beta}\sum_{m_1\cdots m_4}
V_{m_1m_2m_3m_4}b_{m_1\alpha}^*b_{m_2\beta}^*b_{m_3\alpha}
b_{m_4\beta},\nonumber
\end{equation}
with matrix elements
\begin{equation}
V_{m_1\cdots m_4}=\frac{1}{2} \frac{\sqrt{\left(\begin{array}{c}
      N_v\\m_1\end{array}\right)\left(\begin{array}{c}N_v\\m_2
      \end{array}\right)
\left(\begin{array}{c}N_v\\m_3\end{array}\right)
\left(\begin{array}{c}N_v\\m_4\end{array}\right)}}
     {\left(\begin{array}{c}2N_v\\m_3+m_4\end{array}\right)}
     \delta^{m_1+m_2}_{m_3+m_4}.
\end{equation}
This exhibits the dependence on the spatial orbitals. For $c_2\neq
0$, the matrix elements in the additional term consist of
$V_{m_1\cdots m_4}$ multiplied by matrix elements of
$\vec{S}_i\cdot \vec{S}_j$, which depend on the $\alpha_1$,
\ldots, $\alpha_4$ labels of the bosons. These matrix elements can
be found in standard quantum-mechanics texts.

The fact that mean-field configurations break the various
symmetries implies that the minima of the mean field energy form
orbits under the action of these same symmetries.  On a disc, and
at $c_2\neq 0$, one expects and finds that, typically, from a
generic minimum there are 5 flat directions leading to adjacent
minima with equal energy.  These flat directions correspond to the
3 generators of the $SO(3)$ spin symmetry, an overall phase, and
an orbital $O(2)$ rotation.  For spin-independent interactions the
symmetry orbits are generically 10-dimensional.

\begin{figure}
\begin{center}
\epsfig{file=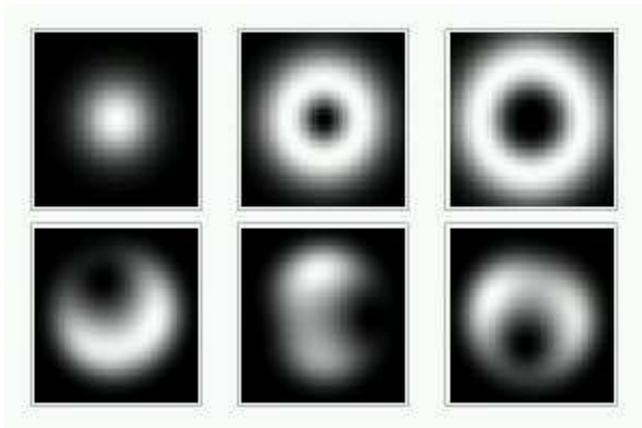,width=8.5cm}
\caption{Two-dimensional density profile of each of the spin
components of two LLL mean field ground state configurations at
$c_2=0$, $L=N$. The upper panels shows the axi-symmetric
spin-vector configuration $(\phi_0(z),\phi_1(z),\phi_2(z))$. The
two configurations share the same distribution of the total
density, and they are related by the $SU(3)$ symmetry. For
$c_2\neq 0$, there are similarly distinct profiles related by
$SO(3)$ symmetry.} \label{btc_dens}
\end{center}
\end{figure}

One convenient quantity to plot is the expectation value $\langle
\vec{S} \rangle$ of the spin, whose length is conserved under
global spin rotations. In special cases, this expectation value is
axi-symmetric; in the more general case it is non-axisymmetric and
the mean field configuration breaks the orbital $O(2)$ symmetry.
Another useful quantity is the three-component condensate
wavefunction (analogous to the familiar spinor for spin-1/2),
which is the expectation value of the field operator, $\langle
\psi_\alpha(z)\rangle=\sum_m b_{m\alpha} \phi_m(z)$ (see
Sec.~\ref{IIA}). It is a vector in the $\alpha=\up$, $0$, $\dow$
basis. From this we can plot the density in each spin component in
position space. This could be accessed experimentally if after
switching off the trap to allow the particle cloud to expand, a
Zeeman term is switched on, which causes the three $\alpha$
components to separate as they expand.

As an example, we plot in figure \ref{btc_dens} the 2D density
profile in each spin component of two different mean field ground
state configurations at $c_2=0$, $L=N$\@.  The top frame shows the
densities for the condensate proportional to
$(\phi_0,\phi_1,\phi_2)$; the lower frame shows a configuration
that is related to this by an $SU(3)$ rotation. The total density
in each of the $m=0$, $1$, $2$ orbitals is an $SU(3)$ invariant,
and it is the same for both configurations shown in figure
\ref{btc_dens}\@. The mean field energy of these configurations is
$E_{MF}={11 \over 48}N^2$, in agreement with order $N^2$ term in
the energy of the exact quantum (BTC) ground state,
eq.~(\ref{Nvenergy}) with $N_v=2$, $p=q=0$ and $n=N/3$.

\begin{figure}
\begin{center}
\unitlength1mm
{\rotatebox{-90}{\epsfig{file=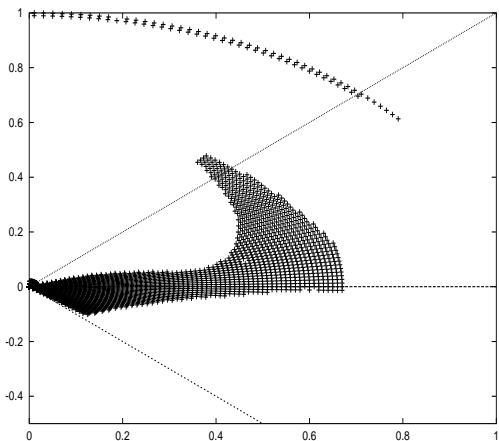,height=7cm,width=6cm}}}
\hskip 2mm \caption{Regions in the $\gamma$, $\ell$ plane in which
only the $m=0$ and $m=1$ orbitals are present in the mean field
ground state on the disc are shaded. The angular coordinate is
$\phi = \arctan{\gamma}$ and $\ell=L/N$ is plotted radially.  In
the shaded strip near $\ell=1$, a ``polar vortex'' forms the
ground state.  The dotted lines mark the $\gamma=\pm 1$ and
$\gamma=0$ directions.} \label{orbs}
\end{center}
\end{figure}

First we consider the disc geometry with $c_2=0$. Carrying out the
mean field minimizations, we find in terms of $\ell=L/N$ that for
$0\leq \ell\leq 2/3$ the number densities $\bra n_m\ket =
\sum_\alpha b^*_{m\alpha}b_{m\alpha}$ in the orbitals of the mean
field ground states behave like (here and in the remainder of this
section, these numbers are normalized so that they sum to 1) $\bra
n_0\ket=1-\ell,~\bra n_1\ket=\ell$ and $\bra n_2\ket=0$\@. For
$\frac{2}{3}\leq\ell\leq 1$ we find $\bra n_0\ket=\frac{1}{3}$,
$\bra n_1\ket=\frac{4}{3}-\ell$ and $\bra
n_2\ket=\ell-\frac{2}{3}$\@. All this is in agreement with the
results derived from the exact quantum ground states in section
\ref{IV}.

For very small interaction ratios $|\gamma|\ll 1$, the total
densities in the orbitals remain the same as for $\gamma=0$, but
there is non-trivial structure in the spin dependence, leading to
spin transitions at critical values of $\ell=L/N$, as we will
describe shortly.

In figure \ref{orbs} we have plotted region II of the phase
diagram, this time with $\ell$ radially. The shaded regions show
where only the first two orbitals ($m=0,1$) are present in the
condensate. One region is a tiny strip near $\ell=1$ for $\gamma
\geq (7+4\sqrt 2)/17\approx 0.75$, where the $(m,\alpha)=(1,0)$
state is occupied by all the bosons. This state can be seen as a
``polar'' vortex, since it has the same spin state as the polar
BEC\@. The other region, centered (roughly) around the $c_2=0$ axis,
contains states in which both the $m=0$ and $m=1$ orbitals are
used.

In the anti-ferromagnetic regime for $\ell\leq 1$ there is a large
area where the $m=3$ orbital requires a non-zero density; in this
area, mean field theory in which only the first three orbitals are
used is not valid.  However, around and at the $SU(3)$-axis and
around the polar vortex as well as in the ferromagnetic regime,
the density in the $m=3$ orbital is very small for $\ell\leq 1$
and can safely be ignored. Besides, if the energy $H_{\omega}$ in
a rotating frame (see eq.~(\ref{H-omega})) is minimized, only the
states which use the first three orbitals $m=0,1,2$ are of
interest for $\ell\leq 1$. [This is with the exception of the
vicinities of the boundaries of region II (see figure
\ref{fig:c-plane}) at $\gamma\to \infty$ and at $\gamma=-1$.]

In the following subsections we present results for the LLL mean field
ground state for $|\gamma|\ll 1$, in both the ferromagnetic and
anti-ferromagnetic regimes, and we discuss the ground states at
$\ell=1$ for general values of $\gamma$.

Our mean field results pertain to the LLL, relevant for the regime
of weak interactions, and they thus differ from the mean field 
solutions of the GP equations \cite{mimaki,mcs}. Nevertheless, 
there is agreement on some of the important features, such as the 
smooth dependence of $L$ on $\omega$ in the ferro regime, and 
the role of the state with a single $\pi$-disclination near $\ell=0.5$ 
in the antiferromagnetic regime \cite{mcs}.
 
\subsection{Anti-ferromagnetic interactions}

We now specify the mean field ground states, given in the form
of a three-component condensate wave function, for small, positive 
$\gamma=+\epsilon$, and for $\ell \leq 1$. As before, the condensate
wave function is a vector in the $\alpha=\up$, $0$, $\dow$ basis.
In the table below we specify the mean occupation numbers
of the four states that we found.

\begin{table}[h!]
\begin{center}
\begin{tabular}{|c|c|c|c|c|c|c|}
\hline
 & $\bra n_{0\up}\ket$ & $\bra n_{0\dow}\ket$ & $\bra n_{10}\ket$ &
 $\bra n_{1\up}\ket$ & $\bra n_{20}\ket$ & $\bra n_{2\up}\ket$\\
\hline

$0\leq \ell\leq \ell^\epsilon_a $ & $\frac{1}{2}(1-\ell)$ & $\frac{1}{2}
(1-\ell)$ & $\ell$
& $0$ & $0$ &$0$\\
\hline

$\ell^\epsilon_a\leq \ell\leq \frac{2}{3}$ & $0$ & $1-\ell$ & $0$ &
$\ell$  & $0$ & 0\\
\hline

$\frac{2}{3}\leq \ell\leq \ell^\epsilon_b $ & $0$  & $\frac{1}{3}
$ & $0$ & $\frac{4}{3}-\ell$ & $\ell-\frac{2}{3}$ & $0$ \\
\hline

$\ell^{\epsilon}_b \leq \ell\leq 1 $ & $0$  & $\frac{1}{3}$ &
$\frac{4}{3}-\ell$ & $0$ & $0$ & $\ell -\frac{2}{3}$\\
\hline

\end{tabular}
\caption{Occupation numbers of the LLL mean field ground state
with small anti-ferromagnetic interaction $\gamma=\epsilon$.}
\label{table1}
\end{center}

\end{table}

Note that the condensates given in this table are specific
representatives of families of condensates that are related by the
$SO(3)_{\rm spin}$ symmetry.  There are two critical values,
$\ell^\epsilon_a=\frac{4}{7}-\frac{\sqrt 2}{7} \approx 0.37$ and
$\ell^\epsilon_b=10-4\sqrt{5}-\frac{4}{3}\sqrt{85-38\sqrt{5}} 
\approx 0.83$, where we see a discontinuous rearrangement of the 
condensate configuration and of $\langle \vec{S} \rangle$. For nonzero 
$\gamma$, these changes in the condensate are continuous; they become 
singular (discontinuous) only as $\gamma\to 0^+$.

\begin{figure}
\begin{center}
\epsfig{file=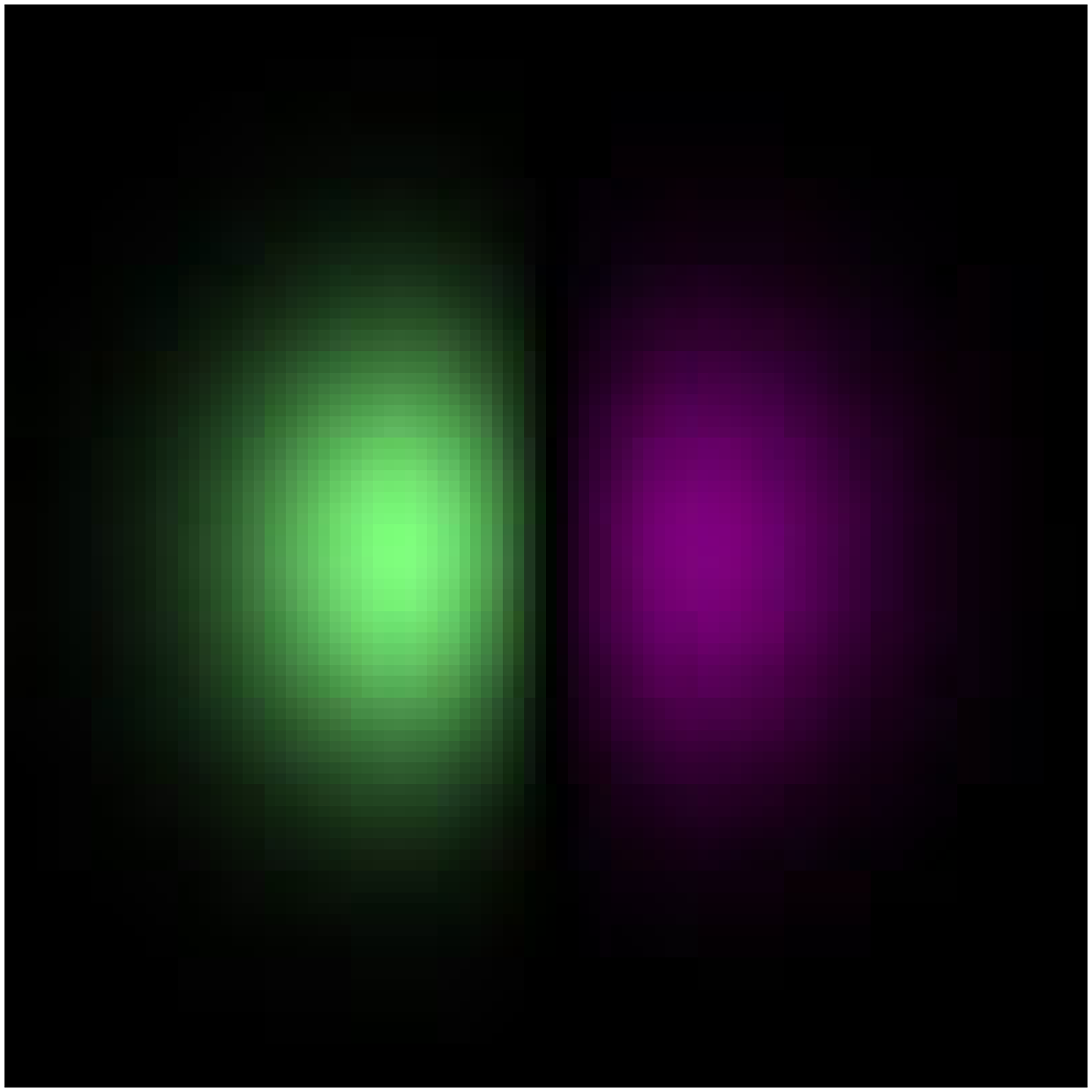,width=35mm}\hfil
\epsfig{file=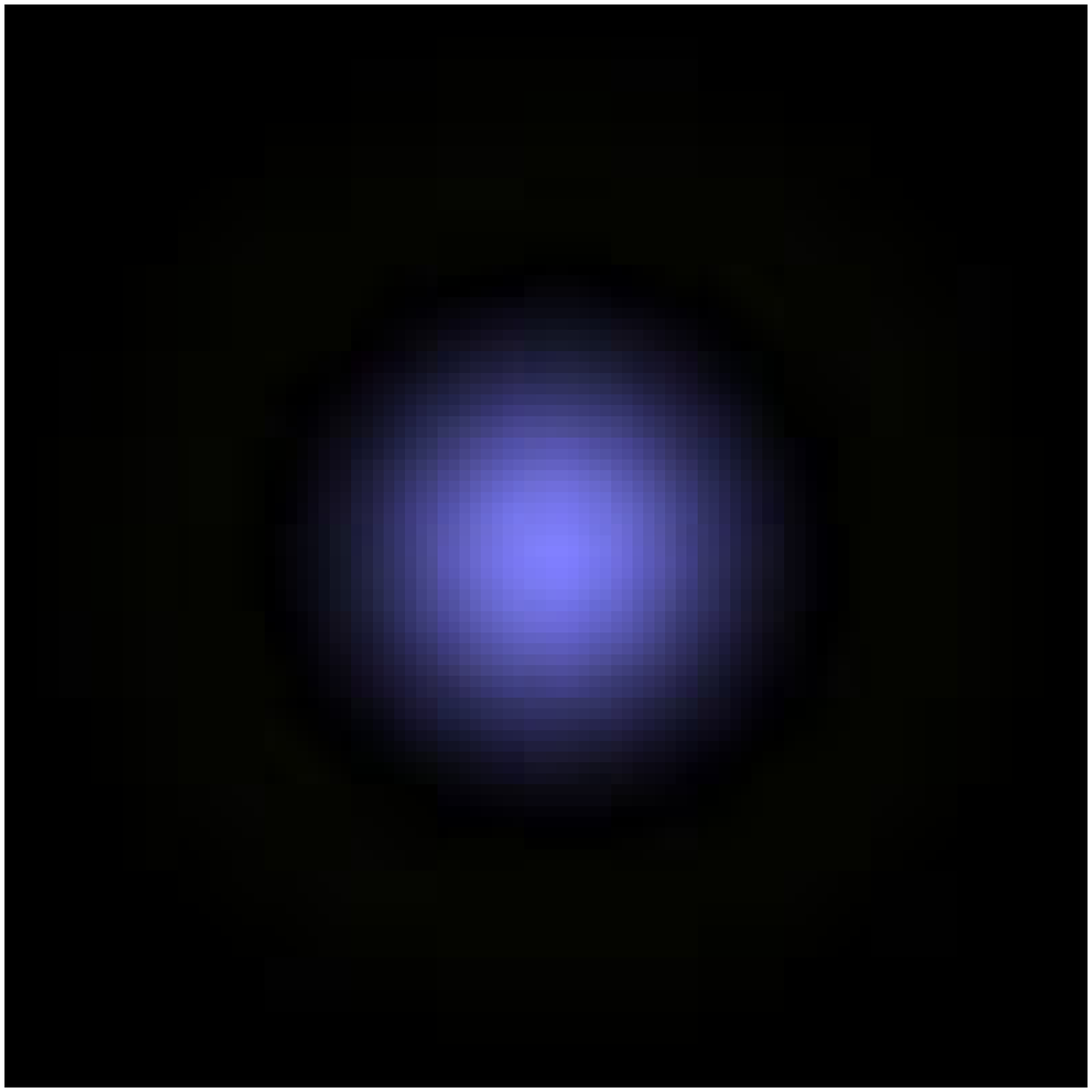,width=35mm}
\end{center}
\caption{Two-dimensional plot of $\bra\vec{S}\ket$ at both
sides of the spin transition at $\ell =\ell_a^\epsilon$. The
intensity codes the length $|\bra\vec{S}\ket|$, while the
color indicates the direction on the spin sphere as in figure
\ref{fig:skyrlat} below. The left and right pictures correspond
to eqs. (\ref{pi_1}) and (\ref{pi_2}), respectively.}
\label{fig:spin_dens}
\end{figure}

For $\ell < \ell^\epsilon_a$, the condensate can be represented 
by $(\frac{\lambda}{\sqrt{2}} \phi_0(z), \eta \phi_1(z), 
\frac{\lambda}{\sqrt{2}} \phi_0(z))$ with
$\lambda=\sqrt{N-L}$, $\eta=\sqrt{L}$\@.  Applying $SO(3)_{\rm
spin}$ rotations, one finds alternative representations such as
$( \frac{1}{\sqrt{2}} [\lambda \phi_0(z) - \eta \phi_1(z)], 0, 
   \frac{1}{\sqrt{2}} [\lambda \phi_0(z) + \eta \phi_1(z)])$\@.  
The $SO(3)_{\rm spin}$-invariant quantity
$|\bra\vec{S}\ket|^2$ is found to be
\begin{equation}
 |\bra\vec{S}\ket|^2 = \frac{N^2}{\pi^2}\ell(1-\ell) 
(z+\bar{z})^2 e^{-2|z|^2} \ .
\label{pi_1}
\end{equation}
The state that emerges at $\ell > \ell^\epsilon_a$ corresponds to
$(\eta \phi_1(z),0,\lambda\phi_0(z))$, leading to
\begin{equation}
 |\bra\vec{S}\ket |^2 = \frac{N^2}{\pi^2}
([1-\ell]- \ell |z|^2)^2 e^{-2|z|^2} \ .
\label{pi_2}
\end{equation}
For $\ell\neq 1/2$, the integrated value of $\langle
\vec{S}\rangle$ for this state is non-zero and there is a
spontaneous magnetization. In figure \ref{fig:spin_dens} a
two-dimensional plot of $\bra\vec{S}\ket$ at both sides of the
spin transition at $\ell =\ell_a^\epsilon$ is shown.
The state at $0 < \ell < \ell_a^\epsilon$ can be viewed as a
configuration of two $\pi$-disclinations off the center of the
trap, while the state in the regime $\ell_a^\epsilon < \ell
< 2/3$ (or possibly even as far as $\ell_b^\epsilon$) can be
understood as a single $\pi$-disclination in the polar state.

The angular momentum for which the $m=2$ orbital is first occupied
in the mean field ground state, $\ell=\frac{2}{3}$, is robust
against small anti-ferromagnetic interactions. 
For $\gamma=+\epsilon$, $2/3 < \ell < \ell^\epsilon_b$, the condensate 
can be represented as $(\tau \phi_1(z),\sigma \phi_2(z),\xi \phi_0(z))$,
with $\xi=\sqrt{\frac{N}{3}}$, $\sigma=\sqrt{L-\frac{2N}{3}}$, 
$\tau=\sqrt{\frac{4N}{3}-L}$, while for $\ell^\epsilon_b< \ell \leq 1$
we have $(-\sigma \phi_2(z),\tau \phi_1(z),\xi \phi_0(z))$.

\begin{figure*}
\begin{center}
\epsfig{file=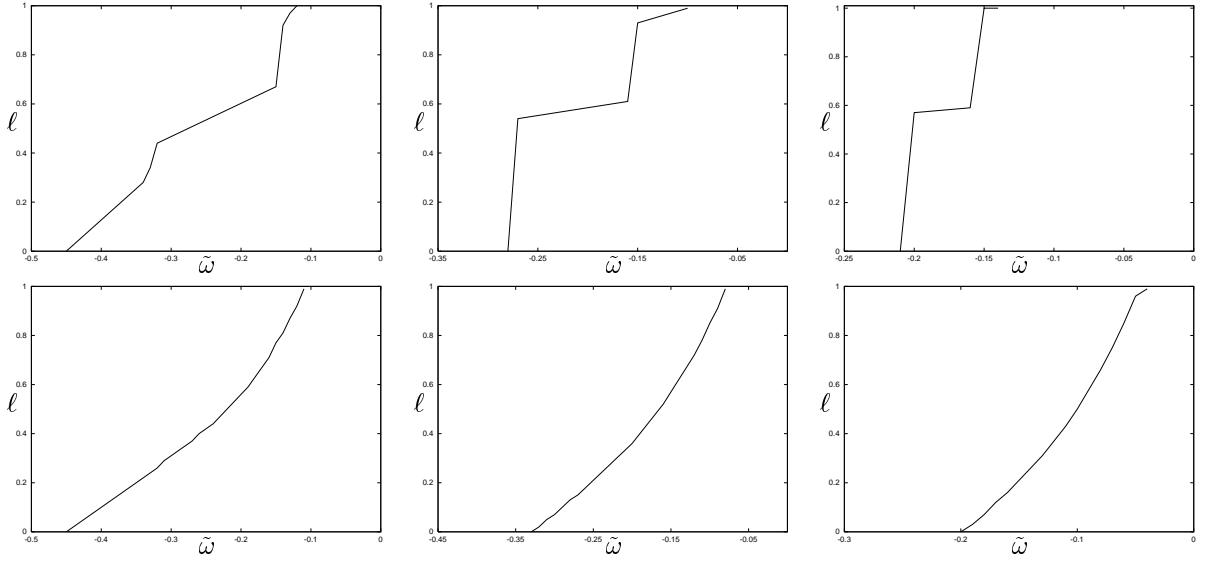}
\end{center}
\caption{The ground state angular momentum per particle $\ell$ on the disc 
as a function of $\tilde{\omega}=(\omega-\omega_0)/c_0N$
for various interaction strengths and slow rotation.  Upper figures:
anti-ferromagnetic regime, $\phi = .1,~.5,~.75$\@.  Lower
figures: ferromagnetic regime, $\phi= -.1,~-.3,~-.5$. 
}
\label{om_beh_GS}
\end{figure*}

In figure \ref{om_beh_GS}, we have depicted the ground state angular 
momentum per particle, $\ell$, as a function of the rotation frequency 
$\omega$ for some positive values of $\gamma$. It is seen that upon
increasing $\gamma$ a semi-plateau (a distinguished part of the curve on 
which the angular momentum increases gradually) develops. 
Upon increasing $\gamma$ further, the semi-plateau becomes flatter and
the width decreases, until for $\gamma$ larger than some critical
value $\gamma_c\approx 1.19$, $\ell(\omega)$ jumps from $\ell=0$ to
an $\ell=1$ plateau at a critical frequency 
$\omega_c$ given by $\omega_0-\omega_c\approx 0.15 
c_0 N$. This is a transition from the non-rotating state to 
the polar vortex, analogous to what occurs in the scalar boson case.

\subsection{Ferromagnetic interactions}

With small negative $\gamma=-\epsilon$ the mean field ground
states for slow rotation are characterized (up to $SO(3)_{\rm spin}$
rotations) by the occupation numbers given in table~\ref{table2}.
Again, we find two spin transitions, the first at
$\ell^{-\epsilon}_a=2-\sqrt{2}\approx 0.59$ and the second at
$\ell^{-\epsilon}_b \approx 0.69$\@.

\begin{table}[ht]
\begin{center}
\begin{tabular}{|c|c|c|c|c|c|}
\hline
 & $\bra n_{00}\ket$ & $\bra n_{0\dow}\ket$ & $\bra n_{10}\ket$ &
 $\bra n_{1\dow}\ket$ & $\bra n_{2\up}\ket$ \\
\hline

$0\leq \ell\leq \ell^{-\epsilon}_a $ & $0$ & $1-\ell$ & $\ell$
& $0$ &$0$\\
\hline

$\ell^{-\epsilon}_a\leq \ell\leq \frac{2}{3}$ & $1-\ell$ & $0$ & $0$ &
$\ell$  & $0$\\
\hline

$\frac{2}{3}\leq \ell\leq \ell^{-\epsilon}_b $ & $\frac{1}{3}
$ & $0$ & $0$ & $\frac{4}{3}-\ell$ & $\ell-\frac{2}{3}$ \\
\hline

$\ell^{-\epsilon}_b \leq \ell\leq 1 $ & $0$  & $\frac{1}{3}$ &
$\frac{4}{3}-\ell$ &
$0$& $\ell -\frac{2}{3}$\\
\hline

\end{tabular}
\caption{Mean occupation numbers of the LLL condensate for small
ferromagnetic interaction $\gamma=-\epsilon$.} \label{table2}
\end{center}

\end{table}

For $\ell<\ell^{-\epsilon}_a$ the condensate can be represented by
$(0,\eta \phi_1(z),\lambda \phi_0(z))$ with $\lambda$ and $\eta$ as 
given above. In this state, the expectation values of the components
of the spin vector take the following form 

\begin{eqnarray}
\bra S_x \ket &=& \frac{N}{\pi\sqrt{2}}\sqrt{\ell(1-\ell)}(z+\bar{z})e^{-|z|^2}
\nonumber \\[2mm]
\bra S_y \ket &=& \frac{N}{\pi\sqrt{2}}\sqrt{\ell(1-\ell)}(-i)(z-\bar{z})e^{-|z|^2}
\nonumber \\[2mm]
\bra S_z \ket &=& \frac{N}{\pi}(1-\ell) e^{-|z|^2} \ .
\end{eqnarray}
The state at $\ell^{-\epsilon}_a < \ell < 2/3$ corresponds to
$(0,\lambda \phi_0(z),\eta\phi_1(z))$, leading to a spin vector 
that vanishes at the center of the disc. 
The spin-textures for $\gamma=-\epsilon$, $\ell<2/3$, can be interpreted as 
half-skyrmions (or merons) (see also section V D below).

For $\gamma=-\epsilon$, $2/3 < \ell < \ell^{-\epsilon}_b$, the condensate 
can be represented as $(\sigma \phi_2(z),\xi \phi_0(z),\tau \phi_1(z))$,
with $\xi$, $\sigma$ and $\tau$ as above,
while for $\ell^{-\epsilon}_b < \ell \leq 1$
we have $(\sigma \phi_2(z),\tau \phi_1(z),\xi \phi_0(z))$.

In the ferromagnetic regime the $\omega$ dependence of the ground state 
angular momentum becomes a smooth curve; see Figure~\ref{om_beh_GS}\@.

\begin{figure}
\begin{center}
\epsfig{file=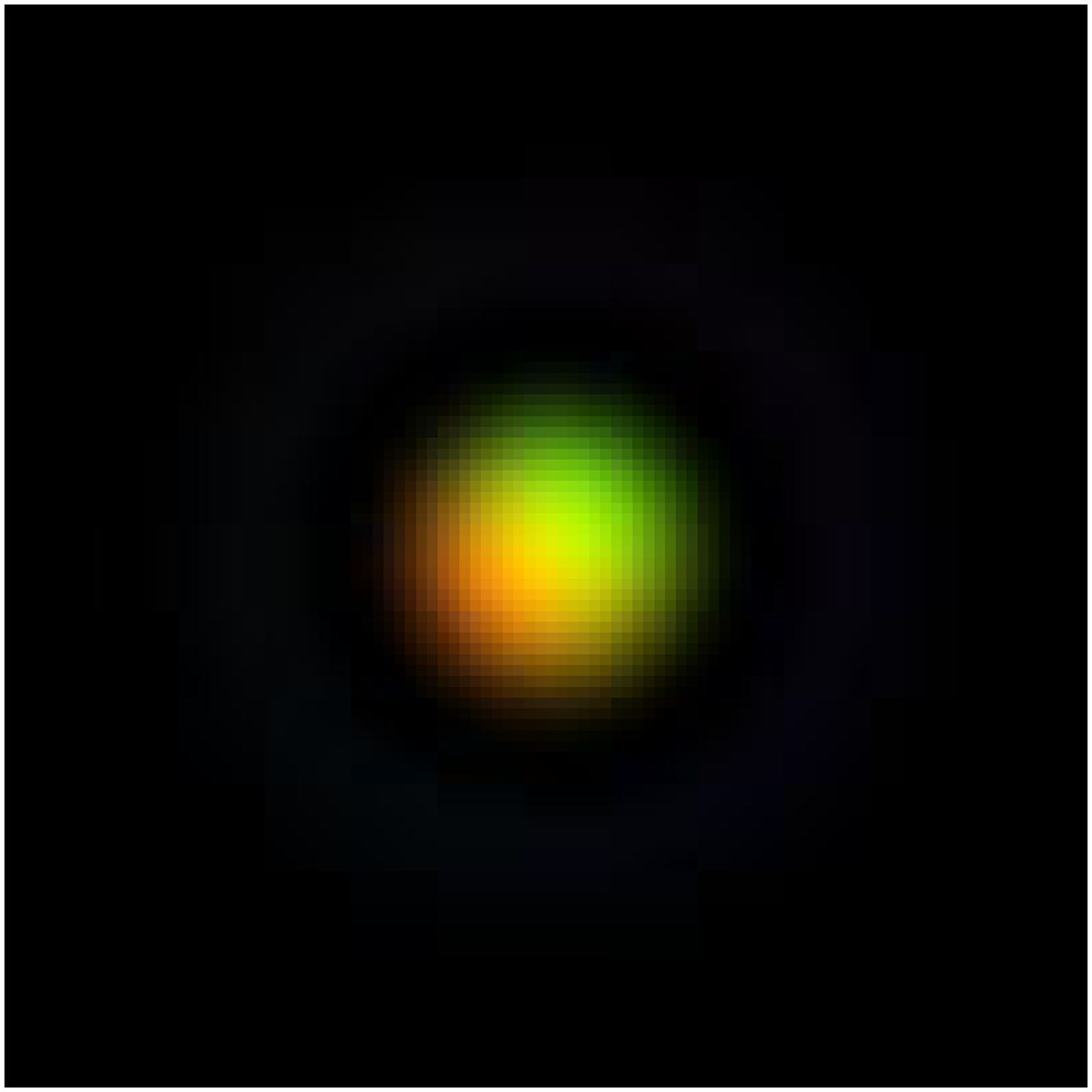,width=35mm}\hfil
\epsfig{file=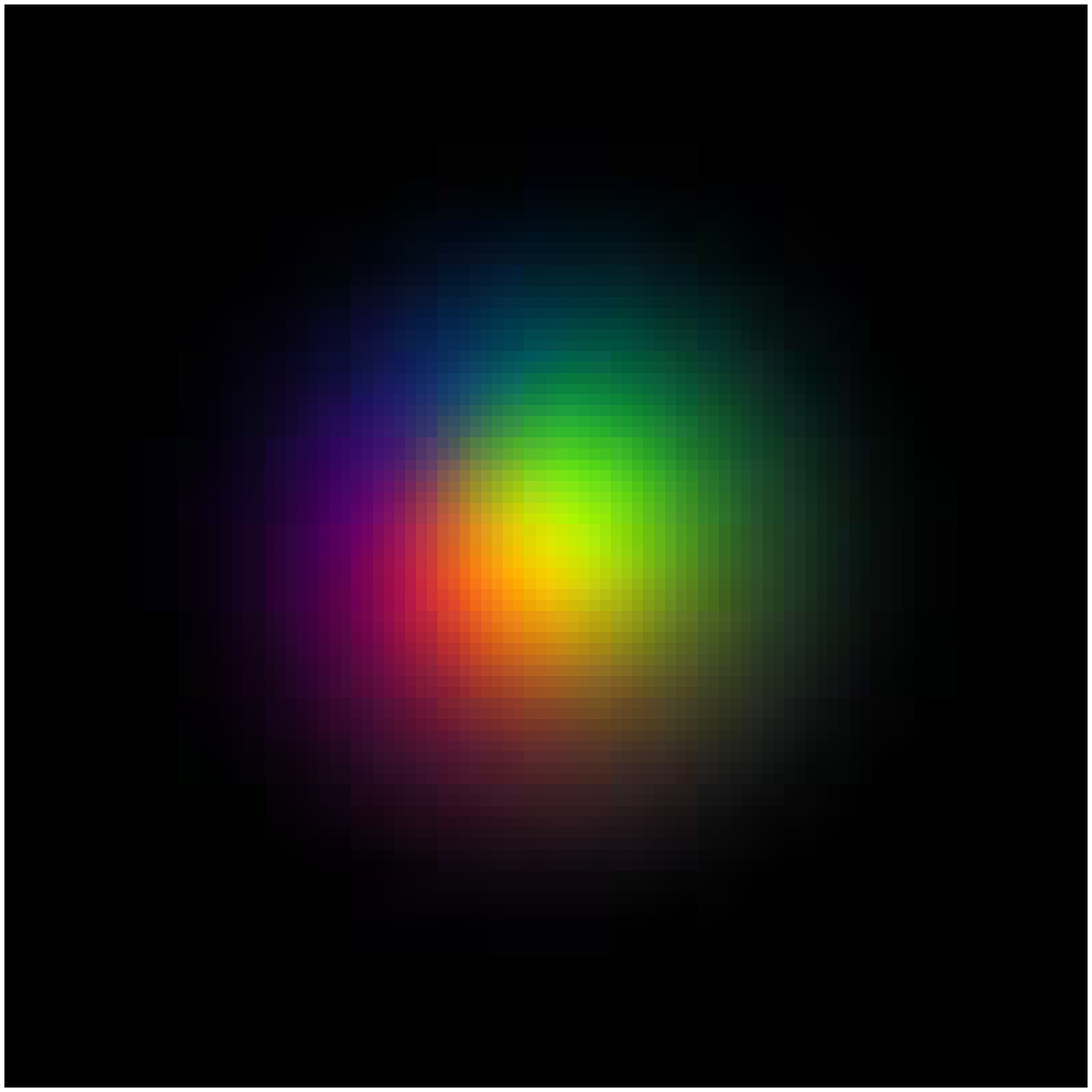,width=35mm}
\end{center}
\caption{The spin texture at $l=1$ for $\gamma=\pm\epsilon$\@.  The left
side is the anti-ferromagnetic case, the right is the ferromagnetic
case. The color coding is as in figure \ref{fig:skyrlat}.}
\label{skyrms}
\end{figure}

\subsection{Mean field configuration at $L=N$}

Assuming that only the first three $m=0,1,2$ orbitals participate
in the ground state, we find that the mean field ground states
at $\ell=1$ take the form 
$(\mp \sigma \phi_2(z),\tau \phi_1(z),\xi \phi_0(z))$,
with $\xi = \sigma = \sqrt{k_{\pm}}$ and 
$\tau = \sqrt{1 - 2 k_{\pm}}$, and with the $(+)-$ sign corresponding 
to (anti-)ferromagnetic interactions. The parameters $k_{\pm}$
depend on $\gamma$ according to
\begin{equation}
k_{\pm}(\gamma) =
\frac{\mp(19+28\sqrt{2})\gamma^2+(42\pm4\sqrt{2})\gamma-3}
     {71\gamma^2+126\gamma-9} \ .
\label{eq:kpm}
\end{equation}

The orbital occupation numbers, given as
\begin{equation}
\bra n_{10}\ket = 1-2k_{\pm}, \quad 
\bra n_{0\dow}\ket =\bra n_{2\up}\ket =k_\pm \ ,
\label{psi_L_N}
\end{equation}
are continuous for $\gamma$ going through $0$, but the 
spin-texture, which is sensitive to the phases in the 
condensate wave function, is not. We find that for
$\gamma=\pm \epsilon$, up to an overall constant,
\begin{eqnarray}
\bra S_x \ket &=& (1 \mp \frac{1}{\sqrt{2}} \bar{z}z)(z+\bar{z})
e^{-|z|^2}
\nonumber \\[2mm]
\bra S_y \ket &=& (1 \mp \frac{1}{\sqrt{2}} \bar{z}z)(-i)(z-\bar{z})
e^{-|z|^2}
\nonumber \\[2mm]
\bra S_z \ket &=& (1 - \frac{1}{2} (\bar{z}z)^2) e^{-|z|^2} \ .
\end{eqnarray}
Note that in the anti-ferromagnetic case, the expectation
value of the spin vector is vanishing on the circle 
$\bar{z}z = \sqrt{2}$, while in the ferromagnetic case we
see a single skyrmion texture with $\bra \vec{S} \ket$
non-vanishing everywhere. Figure \ref{skyrms} shows the spin texture 
at $\ell=1$ for $\gamma=\pm \epsilon$.

From (\ref{eq:kpm}) it is possible to derive the critical
anti-ferromagnetic interaction ratio for which the polar vortex
appears, by simply solving $k_+(\gamma)=0$\@.  The critical value
found then is $\gamma^* =(7+4\sqrt{2})/17\approx 0.75$ (see
figure~\ref{orbs}). If $\gamma$ increases towards $\gamma^*$, the
density in the $m=3$ orbital acquires a small value.  So, strictly
speaking, the states discussed here are not the true mean field 
ground state in the whole intermediate region.
Around $\gamma=0$ and $\gamma=\gamma^*$ however, $\bra n_{3\alpha}
\ket$ is zero and the value of $\gamma^*$ is in agreement with
numerical results.

In the ferromagnetic regime, upon lowering $\gamma$
the parameter $k_{-}$ gradually decreases from
$k_{-}=1/3$ at $\gamma=0$ to $k_{-}=1-\frac{1}{\sqrt{2}}$
at $\gamma=-1$, with the corresponding occupation numbers
given in eq. (\ref{psi_L_N}).

\subsection{The sphere with $N_v=1$, $2$}

It is instructive to perform LLL mean field theory on a system 
of spin-1 bosons in a spherical geometry, with $N_v=1$ or $N_v=2$,
meaning that 2 or 3 orbitals are available to the particles.
To compare with the disc as before, we write these results in terms
of $\ell=(\frac{1}{2}N_v - \tilde{L})/N$. Notice, however, that by 
flattening out the sphere by stereographic projection, the results 
are qualitatively similar to those for the disc when only the
first two or three orbitals are occupied. This is especially true 
for the states at $N_v=2$, $\ell=1$. Even though the
topological classification of textures (see Appendix A) does not
strictly apply to the plane, the form of the spin textures on the
sphere is a useful guide to those in the disc for $\ell\leq 1$.

For the case $N_v=1$ (two orbitals on the sphere), we mention the
following results.  With $0<\gamma\leq\pi/4$ the ground state
configuration is the same as the one we found on the disc for
$\ell<\ell^{\epsilon}_a$. This configuration can be interpreted as two
$\pi$-disclinations at opposite poles of the sphere. For
$\pi/4\leq\gamma<\pi/2$ all bosons occupy the $\alpha=0$ spin component,
forming a polar state with a single vortex.  In the ferromagnetic
regime, with very small $\gamma$ we find the same spin transition as the
one on the disc at $\ell=\ell^{-\epsilon}_a$. With $N_v=1$ this
transition lies at $\ell=1/2$. These configurations can be interpreted
as a half-skyrmion (or meron) in the spin texture. with the spin density
vanishing at one point on the sphere, around which the spin density
winds around the equator in $\vec{S}$-space, passing over one pole at
the opposite end of the sphere.  If the interaction is deformed by
increasing $N_v$ towards $N_v\rightarrow\infty$, the location of the
spin transition is gradually shifted towards
$\ell=\ell^{-\epsilon}_a$\@.  With finite ferromagnetic interaction
($N_v=1$ again), there is a finite region where the core traces a path
over the sphere (from the south pole to the north pole, as $\ell$
increases) and connects the two sides of the transition.  The
interaction energy is clearly independent of $\ell$\@.  This region is
bounded by $\gamma(\ell)=-|\arctan(2\ell-1)|$ for $0\leq\ell\leq 1$.

For $N_v=2$ and $\gamma=0$, the occupation numbers, summed over spin, in
the three available orbitals are given by $\bra n_0\ket=1-\ell,~\bra
n_1\ket=\ell$ and $\bra n_2\ket=0$ for $0\leq\ell\leq\frac{1}{3}$,
followed by $\bra n_0\ket=\frac{5}{6}-\frac{1}{2}\ell, ~\bra n_1\ket
=\frac{1}{3},~\bra n_2\ket = \frac{1}{2}\ell-\frac{1}{6}$ for
$\frac{1}{3}\leq\ell\leq 1$\@.  These mean field results agree with the
exact quantum ground state results obtained in section \ref{IV}.

For $N_v=2$ and small ferromagnetic interactions, $\gamma=-\epsilon$,
the mean occupation numbers in the condensate are given in table
\ref{table3} (up to $SO(3)_{\rm spin}$ and $SO(3)_{\rm orb}$ rotations).
In the trajectory from $\ell=0$ to $\ell=1$ there are no spin
transitions.  The $\ell=1$ state, which has $\bra n_{0\dow}\ket =\bra
n_{10}\ket = \bra n_{2\up}\ket=\frac{1}{3}$, is the mean field ground
state for arbitrary ferromagnetic spin interactions, $0 > \gamma > -1$.
It is a single skyrmion texture with both uniform number density and
magnitude of the spin density, and is discussed further in the Appendix.

\begin{table}[ht]
\begin{center}
\begin{tabular}{|c|c|c|c|}
\hline
 & $\bra n_{0\dow}\ket$ & $\bra n_{10}\ket$ & $\bra n_{2\up}\ket$ \\
\hline

$0\leq \ell\leq \frac{1}{3} $ & $1-\ell$ & $\ell$
& $0$\\
\hline

$\frac{1}{3}\leq\ell\leq 1$ & $\frac{5}{6}-\frac{1}{2}
\ell$ & $\frac{1}{3}$  & $\frac{1}{2}\ell-\frac{1}{6}$\\
\hline

\end{tabular}
\caption{Mean occupation numbers of the LLL mean field ground
state in spherical geometry, $N_v=2$, with small ferromagnetic
interaction $\gamma=-\epsilon$.} \label{table3}
\end{center}

\end{table}

In the case $N_v=2$, and small anti-ferromagnetic
interactions, $\gamma=+\epsilon$, 
for $0\leq\ell\leq\frac{1}{3}$ the mean occupation
numbers per orbital of the condensate are the same as in the
ferromagnetic case, but the spin structure is different.
For $\ell\geq\frac{1}{3}$ the spin expectation values in the
$m=0$ and $2$ orbitals become non-zero (without a discontinuity)
and are not linear functions of $\ell$\@.  Since
$|\bra\vec{S}_{m=1}\ket|^2=0$ and the number density is constant
in the $m=1$ orbital, the spin state describing the bosons in this
orbital can be arranged by an $SO(3)$ rotation to be
$(0,1,0)/\sqrt3$\@.  The vectors representing the bosons in the
$m=0$, $2$ orbitals then are simply constructed. Together with the
previously-mentioned vector they form a mutually orthogonal set
which minimizes $H_n$\@. Provided that the spin-vectors are
properly normalized, the energy can be expressed in terms of one
parameter $\alpha(\ell)$, which is connected to the spin densities
by $\cos[2\alpha(\ell)]=|\bra\vec{S}_0\ket|/\bra n_0\ket
=|\bra\vec{S}_2\ket|/\bra n_2 \ket$\@.  Minimizing the energy with
respect to $\alpha(\ell)$ gives
\begin{equation}
\alpha(\ell)=\arccos\bigg[\sqrt{\frac{1}{2}+\frac{\frac{1}{4}
\sqrt{\lambda(\frac{2}{3}-\lambda)}}{1-4\lambda+6\lambda^2}}\bigg],
\end{equation}
with $\lambda=\frac{1}{2}(\ell-\frac{1}{3})$\@.  The maximum of
the anti-ferromagnetic energy is not dependent on the angular
momentum and lies at $\alpha=\pi/2$\@.  In the ferromagnetic case
this point minimizes the energy, corresponding exactly to the
occupation numbers in table \ref{table3}.

At $\ell=1$, there are solutions with uniform density, with an
unbroken $SO(3)$ subgroup of the $SO(3)_{\rm orb}\times SO(3)_{\rm
spin}$ symmetry, as the limiting case of the previous $\ell<1$
states. This case is also discussed in the Appendix. There are
also solutions in which the orbital distribution in the mean field
configuration of the ground state is not unique. For instance,
among the degenerate states at large $\gamma$ we find the polar
vortex with $\bra n_{10}\ket=1$ and a configuration with $\bra
n_{20}\ket=\bra n_{00}\ket=\frac{1}{2}$.

\subsection{Comparison with finite size exact states}

It is of interest to try to match the mean field states with
ground states found in diagonalization studies, such as those
shown in Figure~\ref{pl} for the disc. Since these have definite
values of the quantum numbers, they can be compared with the mean
field states only by projecting the latter to components with
definite quantum numbers \cite{And}. 
For each $N$ and $L$, the value of the
spin picked out should reflect the form of the interaction, and
should presumably be the maximal value in the ferromagnetic
regime, and the minimal value in the antiferromagnetic. For the
spin-independent case $c_2=0$, the lowest $SU(3)_{\rm spin}$
quantum numbers are favored as ground states.

We will not attempt to identify all the states in Fig.~\ref{pl} in
this way, but only some of the more prominent. We have already
mentioned that the mean field state at $\ell=1$ and $c_2=0$
corresponds to the BTC singlet state. Since the mean field state
has equal mean occupation of $(m,\alpha)=(0,\dow)$, $(1,0)$, and
$(2,\up)$, it does contain a unique singlet component which is
exactly the BTC state. When $c_2$ is turned on, the quantum
numbers remain at $(L,S)=(6,0)$, but the state will be slightly
altered in its details. The corresponding skyrmion spin textures
on the sphere are also discussed in the Appendix.

The BDC multiplet at $L=N/2$ for $c_2=0$ that uses only $m=0$, $1$
also deserves comment. This corresponds in mean field theory to
the $N_v=1$ case discussed above and in the Appendix. When $c_2< 0$,
it becomes a half-skyrmion or meron, which survives for all $-1\leq
\gamma \leq 0$. This meron has no projection to spin 0, and
anyway for this regime maximal spin is expected in the ground
state. Indeed, for $N=6$ the corresponding ($L=3$) state has
$S=3$. The whole regime $L/N \leq 1$ for ferromagnetic
interactions resembles what one expects for skyrmions, that is $L$
(corresponding to $N-\tilde{L}$ on the sphere) decreasing as $S$
increases, as $S=N-L$ \cite{skkr}. For $c_2>0$ and $L=N/2$, the 
lowest-spin part of the BDC state becomes the ground state.

At larger positive $\gamma$, there is a prominent region of
$(L,S)=(6,0)$ in the $N=6$ data. At the largest $\gamma$, we
expect that this can be identified (in the same sense as the
preceding discussion, or as in Ref.\ \cite{ho-yip}) with the polar
vortex state of this section. (In a finite size study, one would
not expect to see a transition from the BTC state at $c_2=0$ to
this polar vortex with the same quantum numbers at large
$\gamma$.) The jump from $L=0$ to $L=6$ expected from the mean
field is seen in Fig.~\ref{pl}. At smaller $\gamma$, a $(3,0)$
region is seen. We speculate that this state corresponds to a
single $\pi$-disclination at the center of the trap (with a second
one at infinity, or the opposite pole on the sphere), and that the
region corresponds to $\ell_a^\epsilon <\ell <2/3$ in the mean
field results. Notice also the prominent semi-plateaus near
$\ell=0.5$ in the plots in Fig.~\ref{om_beh_GS} at larger
$\gamma$.


\section{Vortex and skyrmion lattices}\label{VI}

Upon driving the system faster, multiple skyrmions are induced.  These
are expected to form a lattice and can be well treated in a (quantum
Hall) mean field analysis.  Such an analysis was performed by Kita {\it
et al.} \cite{mfskyr}, who found a range of different lattices for
$c_2>0$, depending on the relative strength to $c_0$ and the rotation.
By including higher Landau levels, they were able to show that some of
these lattices are qualitatively identical at high and low rotation.
Near $c_2=0$, the (scalar) vortex breaks up into three vortices, one for
each spin component, forming a triangular lattice.  For $c_2\geq0.069
c_0$, the vortex splits into two $\pi$-disclinations, which make up a
square (anti-ferromagnetic) lattice.

We have carried out a program, similar to Mueller and
Ho\cite{muellerho}, appropriate for a mean field LLL description of a
multi-component condensate.  The LLL approximation (in the limit
$\omega\to\omega_0$) fixes the vortex lattice spacing to be equal to the
harmonic oscillator length.  Note that this is different from the
Thomas-Fermi regime, where the distance is fixed by the number of
vortices, as the density of bosons is the same as in a non-rotating
trap.

\begin{figure*}
\centerline{\epsfig{file=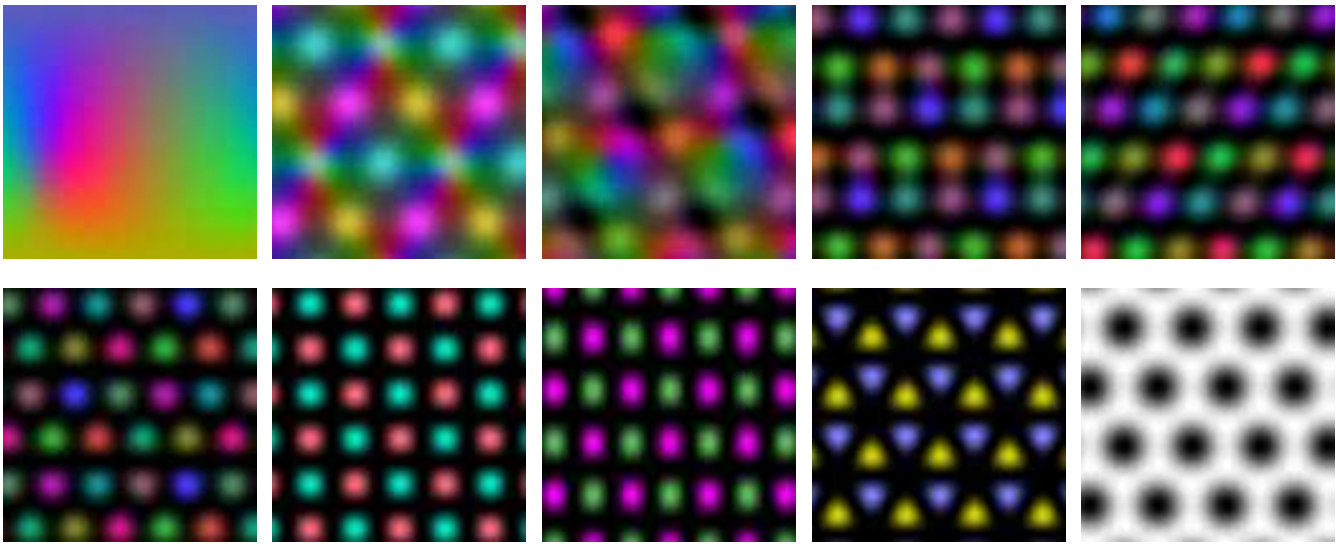,width=\textwidth}}
\caption{The different lattices found in rotating spin-$1$ boson
condensates.  The first picture is the mapping of the spin-sphere
to the colors used.  Top and bottom correspond respectively to the
north and south pole.  The intensity codes $|\langle\vec
S\rangle|$, the size of the spin-vector.  Other pictures are the
spin expectations at different ratios $c_2/c_0$: $\phi=-0.1$, $-0.05$,
$0.01$, $0.016$, $0.04$, $0.1$, $0.54$, $0.7$ and $0.9$.  The last
picture shows the density, as the spin vanishes.}
\label{fig:skyrlat}
\end{figure*}

Under the assumption that the vortices in each spin component form
a Bravais lattice, we can choose the one-particle wavefunctions to
be the torus wavefunctions with $N_v$ flux quanta (typically,
$N_v=1$ or $N_v=2$).  For a scalar condensate, the lattice is
completely specified by the geometry $\tau$ of the torus.  This
wavefunction is periodic up to a gauge transformation, equivalent
to requiring $A'({\bf r'})=A({\bf r})$.

In the case of multi-component condensates, however, more general
boundary conditions are possible.  We only need to demand
\begin{eqnarray}
\psi'({\bf r}+{\bf L}_i)&=&
    e^{i\Lambda_i({\bf r})}\,U_i\,\psi({\bf r}),
\end{eqnarray}
where ${\bf L}_i$ ($i=1$, $2$) define the geometry, and $\Lambda_i$
is the gauge transformation mentioned above.  The matrices $U_1$ and
$U_2$ should commute, $U_1 U_2 U_1^{-1} U_2^{-1}=1$, to obtain
single-valued wavefunctions.

We require that $U_1$ and $U_2$ commute with the Hamiltonian, so
that the energy of a unit cell is well-defined.  For
$\gamma\neq0$, this implies $U_i\in SO(3)_{\rm spin}$\@.  The
common eigenvectors of $U_1$, $U_2$ then have eigenvalues
$(1,e^{i\varphi_1},e^{-i\varphi_1})$ and
$(1,e^{i\varphi_2},e^{-i\varphi_2})$\@.  With an overall $SO(3)$
rotation, we can fix the direction of the vector with eigenvalue
$1$ to be parallel to $\hat z$ in spin space.
With this, the unit cell of the magnetic order (seen in the
spin density which is gauge invariant, for example) is larger than
that of the density, but always contains an integer number of the
latter.

Using this approach, we can confirm a large part of the phase
diagram of Kita {\it et al}\cite{mfskyr}, but we also find
additional phases in the ground states at large $\gamma$. These
are polar phases, for which we use a unit cell with a single
flux quantum. We will use $\phi=\arctan\gamma$ as the parameter. The
minimization procedure
uses a simplex downhill algorithm in the geometry $\tau$ and the
phases $\varphi_1$, $\varphi_2$.  The wavefunction is obtained
from the polynomial free energy by using a conjugate gradient
algorithm, starting from a random point.  The wavefunction in
general is unique up to a phase and a $SO(3)$ rotation along the
$\hat z$-axis.

The phases we obtain, as illustrated in figure \ref{fig:skyrlat}, are as
follows:
\begin{description}
\item[ferro lattice.] A major part of the ferromagnetic phase diagram
    ($-\pi/4\leq\phi\leq-0.08$) is covered by a lattice with $N_v=2$ flux
    quanta in the unit cell. This is the same lattice as one obtains for the
    spin-$\frac12$ bosons with full $SU(2)$ symmetry or, equivalently,
    the quantum Hall ferromagnet with the Land\'e factor $g=0$\@.  If we
    consider the spin-$1$ to be composed of two spin-$1/2$
    particles, then $N_v=2$ for the spin-$1$ bosons corresponds to
    $N_v=1$ for the spin-$\frac12$ particles. This structure is
    related to the $N_v=2$ skyrmions discussed in the Appendix.
\item[skyrmion-vortex lattice.] At $\phi\approx-0.08$, it becomes
	beneficial to include vortices (``merons'').  The unit cell now has
	$N_v=3$, with both a skyrmion and a vortex. Based on direct 
        computations in disc geometry (see below), we expect that this
        phase does not extend to $\phi=0$, but that there are other 
        phases in the weakly ferro regime $-0.02 < \phi< 0$.
\item[triangular vortex lattice.] Exactly at $\phi=0$, the nodes in the
	three components are arranged in a triangular lattice.  This lattice
	can be realized with $N_v=3$ and $\varphi_1=\varphi_2=0$\@.  The
	mean-field components $b_{m\mu}$ ($m=0,1,2$), form a unitary
	($U(3)$) matrix.  This lattice is not shown in figure
	\ref{fig:skyrlat}, as the $SO(3)$-spin is not well defined.
\item[square ladder.] The triangular vortex lattice of the $c_2=0$ case
    is essentially unchanged up to $\phi=0.0143$, being squeezed only.
    However, the $SU(3)$ symmetry is broken. This spin shows a ladder
    structure, where adjacent ladders are shifted by $3/2$ rung-spacings.
\item[canted ladder,] $0.0143\leq\phi\leq0.0193$.  The ladder structure
    stays intact, however the rungs are now canted.
\item[triangular ladder,] $0.0193\leq\phi\leq0.069$.
\item[square $\pi$-disclination,] at $\phi\approx0.069$, there is a
    first order phase transition to the square $\pi$-disclination
    lattice.  Only the $\uparrow$ and $\downarrow$ components are
    present in this lattice.
\item[squeezed $\pi$-discl.,] $0.428\leq\phi\leq0.62$.  The lattice is
    squeezed in one direction and expanded in the other.
\item[triangular $\pi$-discl.,] $0.62\leq\phi\leq0.786$.  At
    $\phi\approx0.62$, there is a first order phase transition to a
    triangular $\pi$-disclination lattice.
\item[polar Abrikosov,] beyond $\phi\approx0.786$, the
    $\pi$-dis\-clinations are unstable and the systems prefers to have
    only one component, such that $\langle\vec S\rangle=0$ everywhere.
    The vortices of this component form an Abrikosov lattice, with
    vanishing density at the cores.
\end{description}
The phases at $\phi>0.428$, and at $\phi<0$ have not been observed before.  
Figure \ref{fig:skyrlat} shows the spin texture in the various lattices,
with colors coding the direction of the spin vector and the intensity
marking its length, so that black regions indicate places where all
components of the spin vector vanish. [For the lattice at $\phi=0.9$,
which is the polar Abrikosov lattice, the spin density vanishes and 
we plotted the particle density instead.] The particle density is finite 
in all lattices except the polar Abrikosov one.

To check whether the Ansatz is sufficiently general in the complete
phase diagram, we have supplemented the above analysis by direct
numerical computations of LLL mean field ground states in a disc
geometry, with $\omega<\omega_0$\@.   Since no periodic structure is
imposed, the lattices form spontaneously.  These computations show that
the torus correctly reproduces the dominant phases such as the square
lattice of $\pi$-disclinations and the skyrmion and skyrmion-vortex
lattices.  In the region $-0.02<\phi<0$ the two geometries showed
different lattice structures, possibly due to finite size effects.  We
leave conclusive results in this region for future work.

At special values of $\varphi_1$, $\varphi_2$, when they are both
of the form $p\pi/q$ ($p$, $q$ integer), it is possible to realize
the lattice by using a larger unit cell and identical phases for
all three spin components.  An example of this is the triangular
lattice at $c_2=0$, where $\varphi_1=-\varphi_2=2\pi/3$.  In this
case, we can realize the same lattice by using a torus with $3$
flux quanta and $\varphi_1=\varphi_2=0$\@.  The other example is the
square $\pi$-disclination lattice, which can be described by using $2$
flux quanta.  This can be compared to the spin-$1/2$
situation\cite{muellerho}, where the lattice at $g_{12}=g_1=g_2$
(unbroken $SU(2)$ symmetry) can equivalently be described by using a
torus with $2$ flux quanta\cite{mfskyr,shankar}.


\section{Quantum Hall liquids}\label{VII}

In the LLL approximation, the filling fraction $\nu$ defines the
average number of bosons in an orbital. Upon increasing the
rotation further, and thus reducing $\nu$, the discreteness of the
occupation numbers becomes important. Mean field theory becomes
less useful due to quantum fluctuations.  At some point, the
condensate is destroyed, and a sequence of quantum fluids takes
over. In the scalar case, the critical $\nu$ was estimated to be
$\nu_c\approx10$ by the Lindemann criterion \cite{CWG} (that is,
the average fluctuation in the position of the vortices equals the
separation between them.) Explicit calculation of small systems
have confirmed this transition and found $\nu_c\approx 6$--$10$\@.
A similar transition will occur in the spin-$1$ case, although we
have not calculated the appropriate $\nu_c$\@. We expect this to
be of the same order of magnitude as in the scalar case. In the
present section we investigate the quantum Hall fluids that appear
within region II of the phase diagram in this regime.

In the extreme limit $\omega\rightarrow\omega_0$ ($\nu\rightarrow
0$), we can analyze the quantum liquids analytically in a part of
the phase diagram, as we can explicitly find the zero-energy
eigenstates of the Hamiltonian. Two of the series we propose, the
$SU(4)_k$ and the $SO(5)_k$ series, have a member of this form for
$k=1$. The third series consists of a generalization of a family
of fractional quantum Hall (QH) states, the hierarchy/composite
fermion states, to spin-1 particles. We present some numerical
results on small sizes, which unfortunately are probably not
conclusive for the nature of the states, due to the restriction to
insufficiently large sizes.

\subsection{$SU(4)_k$ series}

It is straightforward to construct zero-energy states at
$c_2=0$\@. The repulsive contact interaction dictates that the
wavefunction should have a node whenever 2 particles  are on the
same place. Furthermore, two bosons with the same spin should have
a double zero in order to maintain a symmetric wavefunction. In
terms of the components of the wavefunction for given values
$\mu=x$, $y$, $z$ of the spin for each particle, we can write such
a state down \cite{hocluster,paredes,reijnders} (as in Ref.\
\cite{readrezayi1}, we add a tilde to the wavefunction to indicate
that it has to be multiplied by the usual Gaussian factors for the
plane, or the rational factors for the sphere---see Section
\ref{II}):
\begin{eqnarray}
\lefteqn{\tilde\Psi^{2,2,2,1,1,1}(z_1^x,\ldots,z_{N_x}^x;z_1^y,\ldots,z_{N_y}^y;
    z_1^z,\ldots,z_{N_z}^z)\,\,=}\nonumber\\
&&\prod_{\mu=x,y,z}\prod_{i<j}(z_i^\mu-z_j^\mu)^2
  \prod_{\mu'<\mu''}\prod_{i,j}(z_i^{\mu'}-z_j^{\mu''}),
\label{222111}
\end{eqnarray}
where $n_\alpha$ denotes the number of particles with spin
$\alpha$\@.  The lowest angular momentum $L$ for which this state
can be realized is when $N_x=N_y=N_z= N/3$\@. Notice that
$L=N(2N/3-1)$, and the filling factor, which can be defined as
$\nu=\lim_{N\to\infty}N^2/(2L)$, is $\nu=3/4$. This state is a
straightforward generalization of both the Laughlin $\nu=1/2$
state for spinless particles, and the $(2,2,1)$ Halperin
spin-singlet state for spin-1/2 bosons. The full state for
$N_x=N_y=N_z= N/3$ is an $SU(3)_{\rm spin}$ singlet
($(p,q)=(0,0)$), and arguments similar to Laughlin's plasma
mapping show that moreover it has short range spin correlations.
Because the state vanishes whenever the particles are at the same
point, this state is a zero-energy eigenstate for all values of
$c_0$, $c_2$. However, it will only be the ground state (at
$L=N(2N/3-1)$) when $H_{\rm int}$ is positive, that is in the
region $g_0$, $g_2>0$, within region II\@. For larger values of $L$,
there are many more zero-energy eigenstates, so the ground states are
degenerate in the window $g_0$, $g_2>0$. On minimizing $H_\omega$ with
respect to $L$, this implies that the lowest possible filling factor as
$\omega\to\omega_0$ from below is $\nu=3/4$ in this regime (within the
model in Sec.~\ref{II}).

The state in eq.\ (\ref{222111}) is an exact eigenstate, but in
general we are not able to find the exact highly-correlated ground
states of $H_{\rm int}$. Instead, we seek to understand numerical
results, and make predictions for the physics at larger sizes by
using (among other techniques) trial wavefunctions. These states,
which are not generally exact for any know two-body interaction,
serve as paradigms for the phases of matter in the thermodynamic
limit, as they possess interesting (``universal'') properties such
as the quantum numbers and statistics of their excitations that
are robust against small changes in the Hamiltonian, until some
phase boundary is passed (this philosophy has been discussed e.g.
in Ref.\ \cite{readrezayi1}). One way to produce trial
wavefunctions, that is closely connected to their universal
properties, is to use conformal field theory (CFT). We will show
how to obtain wavefunctions from a CFT in somewhat more detail in
the next section. The CFT describing eq.~(\ref{222111}) is
$SU(4)_1$, so following a strategy in Read-Rezayi
\cite{readrezayi1}, and motivated by the analogous results for
scalar bosons \cite{CWG}, we can consider a series, $SU(4)_k$,
where $k=1$, $2$, \ldots. These states have filling factor $\geq
3/4$, and are explicitly $SU(3)_{\rm spin}$ invariant. Hence we
expect them to be relevant near $c_2=0$.

The trial states we consider have wavefunctions, in spin
components (generalizing those in \cite{cappelli,SAvL} to spin-1),
\begin{eqnarray}
\tilde{\Psi}(\{z_i\})&=&{\mathcal S}_{\rm groups}\prod_{\rm
groups}  \tilde{\Psi}^{2,2,2,1,1,1}. \label{eqn:su4k}
\end{eqnarray}
In this construction the $N=3pk$ bosons are first partitioned into
$k$ groups, each with $p$ particles of each spin component $x$,
$y$, $z$.

For each group we write a Halperin $\tilde\Psi^{2,2,2,1,1,1}$
factor, and these are multiplied together. Finally, the
symmetrization operation ${\mathcal S}$ over all ways of dividing
the particles into groups is applied. The angular momentum is
$L=N[2N/(3k)-1]$, and the filling factor (as $N\to\infty$ at fixed
$k$) is therefore $\nu=3k/4$. It happens that if we put $k=N/3$,
the state contains $k$ groups of three bosons each, and the state
is exactly the $L=N$ BTC\@. However, we do not believe this is
particularly significant, as the BTC state is the unique
$SU(3)_{\rm spin}$ singlet state at $L=N$.

The states (\ref{eqn:su4k}) are zero-energy eigenstates of a
Hamiltonian consisting of a $k+1$-body interaction:
\begin{equation}
H_{SU(4)_k}= V\sum_{i_1<\cdots<i_{k+1}}
    \delta(z_{i_1}-z_{i_2})\cdots\delta(z_{i_k}-z_{i_{k+1}}).
\quad
\label{eqn:su4kham}
\end{equation}
The interaction is positive for $V>0$, so all eigenstates have
$E\ge0$\@. This interaction penalizes $k+1$ particles at the same
point. Therefore, zero-energy eigenstates are those in which the
wavefunction vanishes if any $k+1$ coordinates coincide,
regardless of the spins. One can see that the above function has
this property, even before the symmetrization over groups, as for
any $k+1$ particles, at least two must be in the same group,
forcing the function to vanish. For less than $k+1$ particles at
the same point, it does not necessarily vanish. In fact, for each
$k$, (\ref{eqn:su4k}) is the unique zero-energy eigenstate of
$H_{SU(4)_k}$ with lowest angular momentum $L$.

For the same Hamiltonian on a torus, there are again zero energy
states, at least for $N$ divisible by $3k$. For these cases, the
degeneracy of these $SU(4)_k$ ground states is
\begin{equation}
\#_k = \frac16(k+1)(k+2)(k+3) \ .
\label{torus-deg}
\end{equation}
We have verified that this result, which can be inferred from the
CFT connection, is reproduced by exact diagonalization of the
Hamiltonian eq.~(\ref{eqn:su4kham}) on the torus.

Like other incompressible QH states, the phases of matter
exemplified by the trial states (\ref{eqn:su4k}) possess
point-like quasiparticle excitations which may have fractional
particle number (relative to the background density) and/or spin.
The particle number associated with the elementary quasiparticles
can be found once it is understood that, similar to the RR states
\cite{readrezayi1}, the $SU(4)_k$ states are clustered states, in
which particles occur in clusters of $3k$ (in an $SU(3)_{\rm
spin}$-singlet). Then a similar argument to that given in
Sec.~\ref{I} shows that they carry charge $q=\pm 1/4$. They also
have spin 1. The quasiholes, at which there is a deficiency of
particle number, can be studied as zero-energy eigenstates of
$H_{SU(4)_k}$, and fairly explicit trial wavefunctions can be
found using the relation with CFT.

The statistics of quasiparticles in 2D can be defined in terms of
adiabatically dragging them along paths, keeping them
well-separated, to exchange them. For the $SU(4)_k$ states, in the
case $k=1$, the statistics are ``Abelian''; the wavefunction
acquires a phase factor when two particles are exchanged, just as
for the Laughlin states. For $k>1$, however, the statistics
becomes ``non-Abelian'' \cite{mooreread,readrezayi1}. This means
that there is a degeneracy, when the positions of the
quasiparticles are fixed (in general, this is true only when the
quasiparticles are well-separated, though for the $k+1$-body
Hamiltonian and for quasiholes it is exact for any separation). In
terms of the trial functions (\ref{eqn:su4k}) and their
generalization to quasiholes, this degeneracy is caused by the
symmetrization procedure $\mathcal S$, which destroys the
``group'' quantum number of the (quasi-) particles\cite{cappelli}.
The general framework to obtain these degeneracies from CFT has
been worked out and agrees with numerical results for the
Moore-Read, RR, and other states
\cite{mooreread,readrezayi2,GR,ARRS,Ar}. We expect the same
framework to apply here. When the quasiparticles are exchanged
adiabatically, the effect is a matrix operation within these
spaces of degenerate states, described by the braiding matrices of
the corresponding CFT \cite{mooreread}---hence the term
non-Abelian statistics.

\subsection{$SO(5)_k$ series}

Similar to the $SU(4)_k$ case, the $SO(5)_k$ states can be written in the
form:
\begin{eqnarray}
\tilde\Psi^{SO(5)}_k(\{z_i\})&=&{\mathcal
S}_{\rm{groups}}\left[P_{\rm{groups}}
    \tilde{\Psi}^{SO(5)}_{k=1}\right],
\end{eqnarray}
where now
\begin{eqnarray}
\tilde\Psi^{SO(5)}_{k=1}(\{z_i\})&=&{\rm Pf}\left(
\frac{\vec{\zeta}_i\cdot\vec{\zeta}_j}{z_i-z_j}\right)
\prod_{i<j}(z_i-z_j).
\end{eqnarray}
Here the spin states for all the particles are included explicitly
(the product of spin states $\vec{\zeta}_i$ being the tensor
product) ${\rm Pf}\, M_{ij}$ denotes the Pfaffian of an
antisymmetric matrix $M_{ij}$. In the present case the $N=2kp$
particles are partitioned into $k$ groups, with $2p$ particles in
each. The particles in each group form an $SO(3)_{\rm spin}$
singlet. The product over these groups is then symmetrized. The
state as a whole is clearly an $SO(3)_{\rm spin}$ singlet, and has
angular momentum $L=N[N/(2k)-1]$, so the filling factor is
$\nu=k$.

The $k=1$ case, which closely resembles the Moore-Read paired
state \cite{mooreread} but for spin-1 particles, is the exact
ground state of our two-body Hamiltonian $H_{\rm int}$ when
$g_0=0$. That is because, in the state $\tilde{\Psi}^{SO(5)}_1$,
two particles are found at the same point only if they have total
spin 0. Again, the ground state as above is the unique zero-energy
eigenstate at the stated angular momentum, but at larger $L$ there
are many more zero-energy states. So $\nu=1$ is the lowest filling
factor possible at $g_0=0$. This implies that in finite size on
the disc a boundary between ground states with the $L$ values of
the $SU(4)_1$ and $SO(5)_1$ states must run into $\omega=\omega_0$
at $g_0=0$ (this is for $N$ divisible by 6, but there will be
similar statements for other values). Such behavior is seen in
Fig.~\ref{pl}. For $g_0<0$ ($\gamma>1/2$), we do not know the
exact lowest $\nu$ that occurs as $\omega\to\omega_0$ from below.
The $SO(5)_1$ state can be interpreted in terms of BCS
spin-singlet complex--$p$-wave pairing of composite fermions, in
which the Pfaffian represents the pairing in position space
\cite{mooreread,readgreen}.

More generally, for each $k$ there is a Hamiltonian for which the
$SO(5)_k$ states are exact zero-energy eigenstates, again given by
a $k+1$-body interaction:
\begin{eqnarray}
H_{SO(5)_k}&=&V\!\!\!\!\sum_{i_1<\cdots<i_k+1}\!\!\!\!\!
    \delta(z_{i_1}-z_{i_2})\cdots\delta(z_{i_k}-z_{i_{k+1}})\nonumber\\
    &&\quad\quad\quad \times P_{k+1}(i_1,\ldots,
    i_{k+1})\quad.
\end{eqnarray}
This interaction includes a projector
$P_{k+1}(i_1,\ldots,i_{k+1})$ of the spin state of the $k+1$
particles concerned onto total spin $k+1$.

For general $k$, these states can be considered to be built up out
of clusters of $2k$ particles in a spin singlet. From this fact we
can obtain the fractional particle number of the elementary
quasiparticles, $q=\pm 1/2$. Also, the quasiparticle spin is 1/2,
which is fractionalized compared with the spin 1 of the underlying
bosons, and so the number of these quasiparticles must be even.
For $k=1$, there are also excitations with zero particle
number that behave as fermions with spin 1. In this case, the
universal properties may be understood by a simple extension of
the methods of Ref.\ \cite{readgreen} to this case.

Because it is difficult to see through the symmetrization
operation $\mathcal S$, we will provide some details on the
conformal field theory behind these states.  Such a CFT
description allows us to obtain more insight into the topological
properties, such as degeneracies and braiding. For example, to
obtain the degeneracy of ground states on the torus, CFT tells us
that we only need to know the number of non-trivial
representations at level $k$. In the case at hand, this number
turns out to be $\frac12(k+1)(k+2)$\@. Again, we verified this
number using exact diagonalization of the $(k+1)$-body
interaction.

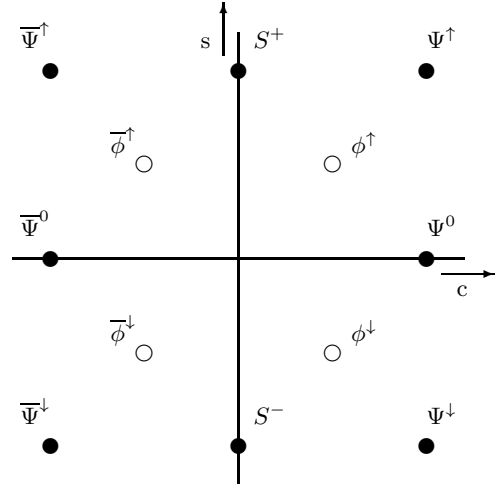
\begin{figure}[ht]
\setlength{\unitlength}{1 cm} \centerline{
\begin{picture}(6,6)(-.5,-.5)
\put(-.5,2.5){\line(1,0){6}} \put(5.2,2.3){\vector(1,0){.7}}
\put(5.4,2.0){c} \put(2.5,-.5){\line(0,1){6}}
\put(2.3,5.2){\vector(0,1){.7}} \put(2.0,5.3){s}
\put(0,0){\circle*{.2}} \put(0,2.5){\circle*{.2}}
\put(0,5){\circle*{.2}} \put(2.5,0){\circle*{.2}}
\put(2.5,5){\circle*{.2}} \put(5,0){\circle*{.2}}
\put(5,2.5){\circle*{.2}} \put(5,5){\circle*{.2}}
\put(1.25,1.25){\circle{.2}} \put(1.25,3.75){\circle{.2}}
\put(3.75,1.25){\circle{.2}} \put(3.75,3.75){\circle{.2}}
\put(5.0,2.8){$\Psi^0$} \put(5.0,5.3){$\Psi^\uparrow$}
\put(5.0,0.3){$\Psi^\downarrow$}
\put(-.4,2.8){$\overline{\Psi}^0$}
\put(-.4,5.3){$\overline\Psi^\uparrow$}
\put(-.4,0.3){$\overline\Psi^\downarrow$}
\put(4,3.9){$\phi^\uparrow$}
\put(.8,1.4){$\overline{\phi}^\downarrow$}
\put(.8,3.9){$\overline\phi^\uparrow$}
\put(4,1.4){$\phi^\downarrow$} \put(2.7,5.3){$S^+$}
\put(2.7,0.3){$S^-$}
\end{picture}}
\caption{Roots and weights of the algebra $SO(5)$\@. The
condensate operators $\Psi$ are associated to roots (filled
symbols) and the fundamental excitations $\phi$ correspond to the
weights of the spinor representation (open symbols).}
\label{fig:so5roots}
\end{figure}

The chiral algebra of the CFT which describes these states is
based on the $SO(5)_k$ affine Kac-Moody algebra. $SO(5)$ is a rank
$2$ Lie algebra, which contains mutually commuting $SO(3)$ and
$U(1)$ Lie subalgebras, which we can identify with the symmetries
under $SO(3)_{\rm spin}$ and number conservation. In these
subalgebras, the generators are respectively $S^+$, $S^-$, $S_z$
and $c$ respectively. According to the CFT-qH
correspondence\cite{mooreread}, we can also obtain the quantum
Hall state wavefunctions as correlators in the chiral part of a
CFT, in which the particles (bosons) should be represented by
fields that have Abelian braiding properties. In the present case
(and the $SU(4)_k$ case is similar), the bosons correspond simply
to a different triplet of current operators of the $SO(5)_k$
affine Kac-Moody algebra. Thus the wavefunction, now in spin
components, can be written
\begin{eqnarray}
\tilde\Psi(\{z_i\})&=&\lim_{z_\infty\rightarrow\infty}z_\infty^{N/k}
    \langle e^{-iN\varphi_c/\sqrt{k}}(z_\infty) \nonumber\\
&&\quad\quad\quad\quad\quad\quad
  J_{\alpha_1}(z_1)\cdots J_{\alpha_N}(z_N)\rangle,
\end{eqnarray}
with $J_\alpha$ ($\alpha=\dow$, $0$, $\up$) an $SO(3)_{\rm spin}$
triplet of currents in the affine Lie algebra, that carry $U(1)$
charge +1. The currents are shown in the $SO(5)$ weight lattice
(fig. \ref{fig:so5roots}), as $\Psi_\uparrow$, $\Psi_0$,
$\Psi_\downarrow$\@. The operator whose position tends to $\infty$
represents a background charge, such that the total $U(1)$ charge
of the operators is zero (as it must be in order that the
correlator be nonzero). The currents can be expressed as
\begin{eqnarray}
J_\alpha(z)&=&\psi_\alpha(z) e^{i\vec\beta_\alpha\cdot\vec\varphi/\sqrt{k}}(z),
\end{eqnarray}
where $\psi_\alpha$ is a parafermion field, of conformal weight
$1-1/k$ for the long roots and $1-1/2k$ for the short roots.  The
vertex operator
$e^{i\vec\beta_\alpha\cdot\vec\varphi/\sqrt{k}}(z)$ contains the
two free boson fields $\varphi_c$ (charge) and $\varphi_s$ (spin),
with $\vec\beta$ the position in the root lattice.  For $J_\alpha$
these are $\vec\beta_\uparrow=(1,1)$, $\vec\beta_0=(1,0)$,
$\vec\beta_\downarrow=(1,-1)$\@.

The parafermions simplify when we specialize to the case $k=1$,
where they reduce to the identity operator for the long roots
($\Psi_\uparrow$ and $\Psi_\downarrow$, in particular) and a
Majorana fermion for the short roots ($\Psi_0$).  The correlator
can then be readily written down, as the correlation functions for
Majorana fermions is well known:
\begin{eqnarray}
\langle\psi(z_1)\cdots\psi(z_n)\rangle&=&{\rm
Pf}\left(\frac1{z_i-z_j}\right).
\end{eqnarray}
This reproduces the $SO(5)_1$ wavefunction, in the same way as for
the spinless Moore-Read state \cite{mooreread}.

We note that the same $SO(5)_k$ algebra was used in another
construction \cite{spchsep}, which was for a system of spin-1/2
particles. The present case differs in that the physical
$SO(3)_{\rm spin}$ symmetry is embedded differently in $SO(5)$,
because of the different spin of the underlying particles.

Wavefunctions for zero-energy states containing quasiholes can
also be written down as chiral correlators, which now contain
vertex operators for primary fields of the $SO(5)_k$ algebra that
represent the quasiholes. For the $k=1$ case, these contain (in
the scalar field plus Majorana fermion language) a spin field for
the Majorana fermion, and give rise to quasihole wavefunctions
analogous to those for the Moore-Read state
\cite{mooreread,readrezayi2}.

\subsection{Composite fermions}

Alternative QH states to the rather exotic series in the previous
two subsections can be constructed by applying conventional
methods to spin-1 bosons. One such approach, as in the case of
scalar bosons, is to map the bosons onto composite fermions, by
attaching (say) one vortex to each boson. These fermions see a
reduced effective magnetic field, and one can construct an
incompressible state when an integer number of Landau levels in
the effective magnetic field is filled with all three components.
This construction gives states with filling factors
$\nu=3p/(3p\pm1)$, which are $SU(3)_{\rm spin}$ singlets.

One can interpret the Moore-Read state as the $p$-wave pairing of
composite fermions \cite{mooreread}.  In this case, $p$-wave 
$SO(3)$-singlet pairing is possible (in contrast to the spin-$1/2$ case) 
and indeed, we have seen that the $SO(5)_1$ state can be interpreted
this way. In the $SU(3)$ symmetric case at $\nu=1$, no 2-particle
$SU(3)$-singlet pairing is possible and there are two options for
the system.  One is to form a Fermi liquid, the other is to
spontaneously break the symmetry and form $(p,q)=(2,0)$ pairs.
Note that this last possibility includes the $SO(3)$ singlet and
thus can be continuously connected to the $SO(5)_1$ state.

\subsection{Vortex lattices without polar order, and nematic QH liquids}

The earlier discussion of QH liquid states focused on singlets
under $SO(3)_{\rm spin}$, with short range spin correlations. It
is interesting to wonder also if QH liquids with some form of spin
ordering could occur. One possibility would be a ferromagnetic QH
liquid. Such states can be easily written down, by using any
wavefunction for a QH state of spinless bosons, with all the boson
spins in the $\alpha=\up$ state (or a global spin rotation of
this). One might expect these to occur in the ferromagnetic
($c_2<0$) part of region II, but in fact we see no sign of them:
leaving aside the skyrmion textures in the BEC at low $L$, at
larger $L$ all ground states are spin singlets. We note that for
spin-1/2 electrons, spin-polarized states can occur, e.g.\ at
$\nu=1$, even for spin-independent interaction, due to exchange
effects. However, the exchange effects are presumably different
for bosons.

A more feasible-looking possibility is QH states with polar spin
order, perhaps in the antiferromagnetic region $c_2>0$. In the
regime at large $\gamma$ where mean-field theory predicts the
Abrikosov vortex lattice, the spin-order is polar. In the polar
state, the vector condensate can be written as
$\langle\psi_\mu\rangle=e^{i\varphi}n_\mu$, with $\varphi$ the
phase and $\hat n$ a real vector. In Abrikosov lattice, the
magnitude of the vector $\hat n$ and the phase $\varphi$ vary to
give a triangular lattice of vortices. We can now imagine that
quantum fluctuations destroy either part of the order (restoring
either the phase or the spin-rotation symmetry) without the other.
When the $U(1)$ and translational symmetry that are broken in the
vortex lattice are restored, the ground state is a QH fluid. For
large quantum fluctuations one might expect that the QH liquid has
restored $SU(2)_{\rm spin}$ symmetry. However, the two transitions
at which these symmetries are restored are independent, and the
transitions could in principle occur in either sequence as we go
to smaller $\nu$. The intermediate phase in which spin symmetry is
restored but not the phase would be a vortex lattice in a boson
paired state, a BEC of boson pairs. This would be characterized by
having a nonzero expectation value of $\sum_\mu \psi_\mu(z)
\psi_\mu(z)$, which is invariant under $\psi\to -\psi$. (Such a
vortex lattice would also be possible with vortices containing a
half-unit of vorticity each, instead of integers as we assume
otherwise, and this might be reached by restoring symmetry in the
$\pi$-disclination lattice state.)

The other possible sequence of transitions would be that in which
the intermediate phase is a QH liquid with restored translational
and phase symmetry, but still has the polar order, which breaks
$SU(2)_{\rm spin}$. The single-boson expectation $\langle
\psi_\mu(z)\rangle$ would be zero, but if we look at the composite
operator $\psi^\dagger_\mu(z) \psi_{\mu'}(z)$, this can have an
expectation value.  This matrix has a trace equal to the density,
which is uniform by assumption. The traceless Hermitian matrix
obtained by subtracting off the trace contains antisymmetric and
symmetric parts. The (imaginary) antisymmetric part corresponds to
a spin-1 irreducible tensor that is simply the spin density, which
is assumed to be zero here. The (real) symmetric part corresponds
to a spin-2 irreducible tensor. This is the order parameter of a
polar or nematic state, which represents a vector $\hat n$, but is
invariant under $\hat n\to -\hat n$, so it parametrizes
$S^2/Z_2=RP^2$\@. Trial wavefunctions for these nematic quantum
Hall states can be written down as those for scalar bosons, times
a spin state such as $\alpha=0$ for all bosons, or as a
spin-rotation of this.

It would not be surprising if such nematic QH liquids occurred in
the phase diagram at large $\gamma$, now that the corresponding
(polar Abrikosov) vortex lattices are known to be present. In
finite size, the ground state would always be low spin
\cite{And}, $S=0$ or
$1$, and there need be no transition separating it from a state at
the same $L$, $S$ with short-range correlations, such as the
$SO(5)_1$ state at $g_0=0$. Thus the appearance of such nematic
order in a QH fluid of spin-1 bosons in the thermodynamic limit
cannot yet be ruled out.

\subsection{Numerical results}

To examine how well the proposed states describe the true ground
states, we have performed exact diagonalization of small systems.
In the regime $L\leq N$, we have used both the disc and sphere
geometries.  As we have seen, these results differ somewhat. But
when looking at fast rotation, however, where the filling factor
is of order 1, the system is spread out into a pancake. It makes
sense to focus attention on the interior of the disc and avoid
edge effects. This can be done by using an edgeless geometry such
as the sphere or torus. Here we will be interested in the ground
states in which (unlike the work earlier in this paper) we find
the ground states without constraining $L$. QH liquid ground
states will usually then show up as $\tilde{L}=0$ states. At
finite sizes on the sphere, such ground states that form a
sequence of sizes tending to a particular filling factor $\nu$ in
the thermodynamic limit lie on a sequence of the form $N_v=N/\nu-s$
\cite{haldane} . Here $s$ is known as the shift, and its
appearance is connected with the coupling of the particles to the
curvature of the sphere. The value of $s$ depends on the liquid
state, not only on $\nu$. $N_v$ can be obtained from the angular
momentum on the disc as $L=NN_v/2$ (all states have $\tilde L=0$).
That is, when the states in the $\tilde\Psi$ notation are written
for the sphere, one can take $N_v$ as small as possible, so that
$N_v$ equals the highest power of any $z_i$ appearing in the
wavefunction (see Sec.~\ref{II}). For QH ground states, this will
usually mean that $\tilde{L}=0$. For the $SU(4)_k$ ground states,
we have $N_v=4N/(3k)-2$, while for the $SO(5)_k$ ground states
$N_v=N/k-2$. We will compare the results of numerical solution for
the ground states with these series of trial states.

The $SU(4)_k$ states with $k=N/3$ are the exact ground states for
$c_2=0$ at $N_v=2$ on the sphere. It is not surprising that they
are eigenstates, because they are the only $SU(3)$ singlet states,
as in the case of $L=N$ for the disc.

As a further test for the $SU(4)_k$ states, we have looked at
sizes $(N,N_v)$ at which such a ground state could lie for $k>1$.
Since $N$ must be divisible by $3k$, such sizes increase rapidly
even for $k=2$. The next case after the trivial BTC for $k=2$ is
$N=12$, $N_v=6$. Here it turns out that the overlap-squared of the
exact ground state for $c_2=0$ with the trial state is \cite{Re}
$|\langle SU(4)_2|G.S.\rangle|^2 = 0.915226$.

The $SO(5)_1$ state was shown to be an exact ground state for
$c_0>0$, $g_0=0$. The higher members of this series, however, had
a vanishing overlap with the ground states throughout the phase
diagram.

As a further test of our proposed wavefunctions, we have
performed calculations for torus geometries. On the torus, one
simply has $\nu=N/N_v$ for the finite size sequence of ground
states that tend to a fluid of filling factor $\nu$ in the
thermodynamic limit. We saw in the mean field analysis of the
skyrmion lattice, that the lattice can only be observed when the
number of flux quanta is a multiple of three. However, at low
filling factors, we expect to see quantum Hall states at
$\nu=3k/4$\@. To be able to observe these, $N_v$ has to be a
multiple of $4$\@. Unfortunately, this implies torus sizes which
are too large to observe both the quantum liquids and the skyrmion
lattice.

To see if the proposed wave-functions are good candidates, we are
therefore forced to look at tori which frustrate the mean field
skyrmion lattice.  The cases we considered are $N_v=3$, $4$, $6$.
For $N_v=4$, we find that the ground states are exactly given by
the $SU(4)_k$
series. However, as for the BTC states on the sphere, this is due
to the fact that the trial ground states, which are $SU(3)_{\rm
spin}$ singlet states, span the space of all $SU(3)_{\rm spin}$
singlets on the torus, which has dimension equal to
eq.~(\ref{torus-deg}), the degeneracy of $SU(4)_k$ torus ground
states. Clearly this must be independent of the geometry of the
torus (described by $\tau$), and we verified this in some cases.

\begin{figure}
\epsfig{file=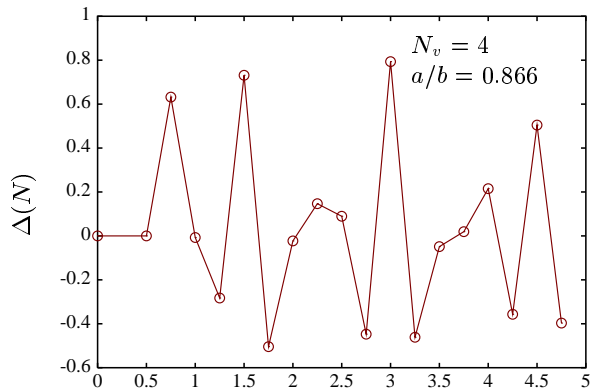,width=8.5cm}
\caption{Particle-hole excitation gap $\Delta(N)$ versus $\nu$,
for $N_v=4$, in a rectangular geometry, $a/b=\sqrt{3}/2$. The
peaks can be interpreted as an indication of incompressibility
of the corresponding states.  For $N=3$,$6$,$9$ we
verified that the ground states, which are degenerate, are exactly
the $SU(4)_k$ quantum Hall trial states with $k=1$, $2$, $3$,
respectively.} \label{fig:torus4}
\end{figure}

In fig.~\ref{fig:torus4} we have plotted the particle-hole
excitation gap
\begin{equation}
\Delta(N) = N\left(\frac{E(N-1)}{N-1}+\frac{E(N+1)}{N+1}-2\frac{E(N)}N\right),
\end{equation}
where $E(N)$ is the ground state energy for $N$ particles, for
$N_v=4$. In the thermodynamic limit, this quantity will exhibit
upward peaks at filling factors that corrrespond to incompressible
states. We performed the diagonalization for total $SO(3)$ spin
$S=0$, $1$, $2$.

\begin{figure}
\epsfig{file=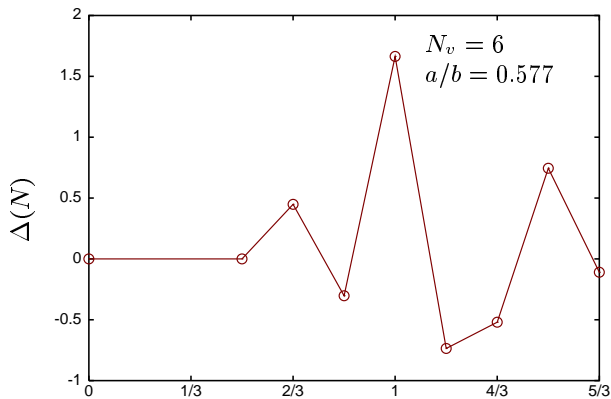,width=8.5cm}
\caption{Particle-hole excitation gap $\Delta(N)$ versus $\nu$,
for $N_v=6$, in a rectangular geometry, $a/b=\sqrt{3}/3$.}
\label{fig:torus6}
\end{figure}

For $N_v=6$ (figure \ref{fig:torus6}), we focussed on the state at
$\nu=3/2$, $N=9$, which corresponds to $k=2$.  We have calculated
the overlap-squared with the $SU(4)_2$ state to be 0.939804.
Another feature in the $N_v=6$ plot is the state at $\nu=1$
($N=6$).  This could be a precursor to a paired composite fermion
state; however, the overlap with the $SO(5)_1$ state was small.

\subsection{The boundaries of region II}
The behavior at the phase boundaries I--A/II and I--B/II (see
figure \ref{fig:c-plane}) deserves special attention.  At the
boundary I--B/II, where $\gamma=c_2/c_0=-1$, the Hamiltonian
simplifies as $g_2=0$ and only the contact interaction which
projects onto the spin-singlet channel remains.  As a result of
this, large degeneracies occur. For example, all fully-polarized
states ($S=N$) have zero energy.  We have not obtained analytic
expressions for these degeneracies. That they are not due to the
specific geometry was observed on the torus.  The zero energy
states are not sensitive to changes in the geometry.  As examples
of these degeneracies, we have in figure \ref{fig:c2c0} tabulated
the $\tilde{L}$, $S$ quantum numbers of the zero energy states for
$N=6$, $N_v=2$ and for  $N=5$, $N_v=3$ on the sphere, and for
$N=6$ particles in the disc geometry.

\begin{figure}[!ht]
\vskip 6mm
\begin{tabular}[t]{c|ccccccc}
$S\backslash \tilde{L}$&0&1&2&3&4&5&6\\\hline
0&1\\
1&&1\\
2&&&1\\
3&&&&1\\
4&1&&1&&1\\
5&&1&1&1&1&1\\
6&1&&1&&1&&1
\end{tabular}
\hskip5mm
\begin{tabular}[t]{c|cccccccc}
$S\backslash \tilde{L}$&$\frac12$&$\frac32$&$\frac52$&$\frac72$&$\frac92$&%
    $\frac{11}2$&$\frac{13}2$&$\frac{15}2$\\\hline
0& 1&&1\\
1& 1&2&1&1\\
2& 1&2&2&1&1\\
3& 1&2&2&2&1&1\\
4& 1&2&2&2&2&1&1\\
5& &1&1&1&1&1&&1\\
\end{tabular}
\vskip2mm
\begin{tabular}[t]{c|ccccccccc}
$S\backslash L$&$0$&$1$&$2$&$3$&$4$&$5$&$6$&\ldots\\\hline
0& &&&&&&1&\ldots\\
1& &&&&&1&1&\ldots\\
2& &&&&1&1&3&\ldots\\
3& &&&1&1&3&5&\ldots\\
4& &&1&1&3&4&8&\ldots\\
5& &1&1&2&3&5&6&\ldots\\
6& 1&&1&1&2&2&4&\ldots\\
\end{tabular}

\caption{Degeneracies of zero-energy ground states on the sphere
          at $\gamma=-1$ for $N=6$, $N_v=2$ (top left), $N=5$, $N_v=3$ (top right)
         and $N=6$, $N_v=\infty$.
         All multiplicities refer to highest weight states of
         the orbital $SO(3)$ symmetry.}
\label{fig:c2c0}
\end{figure}

\section{Conclusion}

In this paper we have studied the phase diagram of spin-1 bosons
in a rotating trap, within the LLL approximation, using a variety
of techniques (numerical diagonalization, mean field theory, and
analytical constructions). We concentrated on certain regimes.
These were (i) low rotation, such that the angular momentum $L$ is
less than or equal to the particle number $N$, where the system is
beginning to contain some vorticity; (ii) higher rotation, where
the bulk of the fluid accommodates vorticity and is occupied by a
lattice of (possibly coreless) vortices, which we considered as
infinite periodic structures; (iii) the quantum Hall regime, in
which the vortex lattices are replaced by
translationally-invariant quantum fluids, which we considered in
edgeless geometries (unfortunately, for spin-1 bosons the finite
size restrictions are here very severe). The transition regions
between these regimes, namely that in which the system contains a
small number (larger than two) of vortices, and that at the
critical filling factor at which the vortex lattices are replaced
by the quantum liquids, were not considered. The results show a
rich variety of phases as the interaction parameters, especially
the ratio of the coefficients of spin-dependent and
spin-independent interaction terms, are varied. The results
obtained here, especially those at lower rotations which should be
more easily accessible, should motivate further experiments to
rotate spin-1 bosons with  unbroken spin-rotation symmetry.

\subsection{Acknowledgements}
We thank Ed Rezayi for collaboration in an early stage of
this study and for discussions; we acknowledge discussions with
T.-L.~Ho, E.J.~Mueller, R.~Shankar.
This research was supported by the Netherlands Organisation for
Scientific Research, NWO, and the Foundation FOM of the Netherlands 
(J.W.R., F.J.M.v.L. and K.S.), and by the NSF under grant no.\ 
DMR-02-42949 (N.R.).

\begin{appendix}

\section{Classification of topological defects}

Here we continue the discussion of the topological classification
of defects or excitations.

In Sec.\ \ref{I} we explained the appearance of two types of
ordered BECs for spin-1 bosons. In these the order is constant in
space. More generally, a condensate will prefer to have the same
type of order locally at almost all points in order to lower the
energy, but there are types of excitations in which the order as
described by a point in the order parameter manifold can vary in
space. Excitations of this type that are stable under continuous
deformations of the order (known as topological defects) can be
classified by methods from topology. One type is those in which
the order breaks down (possible, the density goes to zero) at a
point in space (we consider two space dimensions here). These are
classified by the fundamental or first homotopy group $\pi_1$ of
the manifold. Such objects exist for example in the case of a
one-component (scalar) condensate, where the order parameter
manifold is the circle $S^1$. For vector condensates, an example
is the polar state, for which $\pi_1(S^1\times S^2/{\mathbb
Z}_2)={\mathbb Z}$, the group of integers. Because these involve
the phase of the condensate (related to the $S^1$ part of the
manifold) winding as one encircles the singular point, these
objects in both examples are vortices that are relevant when
vorticity is forced into the system. In the polar state, the
vortices of the smallest vorticity carry a half unit of vorticity
(in the usual units), because of the ${\mathbb Z}_2$ divided out,
and are also referred to as $\pi$-disclinations. For the case of
the ferromagnetic state, $\pi_1(SO(3))={\mathbb Z}_2$, so that two
nontrivial vortices can annihilate one another \cite{Ho}.

The other main type of topological defect is sometimes called a
coreless vortex or skyrmion. In these the order exists and varies
within the order parameter manifold everywhere in space. The
topological classification requires identifying points at
infinity, as if space were a sphere. For trivial boundary
conditions (those that allow the constant ordered ground state),
the topological defects are classified by the second homotopy
group $\pi_2$ of the manifold. This is $\mathbb Z$ in the polar
case, and trivial in the ferromagnetic case. However, if we wish
to classify the vortices that carry the nonzero vorticity in a
rotating condensate, then the boundary conditions must be modified
to allow a nonzero vorticity on the sphere (this modification has
of the form of the Dirac string familiar for a magnetic monopole).
In the polar case, the presence of net vorticity may force
vortices into the ground state. Whether the ground state contains
$\pi$-disclinations, or vortices of larger vorticity each, depends
on the detailed energetics, but one should note that as the total
vorticity is an integer $N_v$, the number of $\pi$-disclinations
(more generally, the number of vortices that carry half-integer
vorticity) must be even. On the other hand, the defects above that
are classified by $\pi_2$ carry no vorticity. For the
ferromagnetic case, there are coreless vortices or skyrmions which
are topologically-nontrivial textures in the order and carry
nonzero vorticity quantized in integers. We will discuss these
configurations in more detail below.

We now come to the borderline case in which the interaction is
spin-independent ($SU(3)_{\rm spin}$ invariant), $c_2=0$. This is
a useful starting point for small $\gamma$ also. The discussion is
easily generalized to the case of an $n$-component order parameter
with an $SU(n)$ invariant interaction.  In such a case, the target
space is described by a complex $n$-component vector of unit
length, proportional to the expectation value of the boson
operator, which lies in $U(n)/U(n-1)\equiv S^{2n-1}$\@. For $n>1$,
$\pi_1=\pi_2=0$, so this space has neither point-singular vortices
nor skyrmions without vorticity. For $n=1$, there are the
well-known point-singular vortices with one unit of vorticity
each, and for $n>1$ the presence of nonzero vorticity on the
sphere induces skyrmions with integer vorticity in the condensate,
as we will now see.

In more detail, the Bose condensate on a sphere with vorticity
present is described by the expectation value of the field
operator $\langle \psi_\mu(z)\rangle$ as $z$ ranges over the
sphere. Thus the possible condensates form the space of sections
of a complex vector bundle. Such bundles are classified
topologically (for each number of components $n>0$) by their first
Chern class, which is simply the (integer) number of vortices
$N_v$ that we have been using, or the number of flux quanta in the
monopole at the center of the sphere \cite{haldane}. Restricting
the bosons to the LLL means considering only the ``holomorphic''
sections of the same bundles. The LLL mean field theory performed
in Section~\ref{V} simply finds such holomorphic sections of
lowest mean-field energy. Condensates in which there are $N_v$
vortices at which the density (magnitude squared of the
condensate) vanishes always exist; simply take the condensate
entirely in one spin component. The question we will pursue here
is the existence of coreless vortices or skyrmions with non-zero
vorticity, in which the condensate is nonzero at all points on the
sphere. These correspond to nonsingular configurations of the
order as described above.

The nicest configurations of all, which serve to illustrate the
most elementary skyrmions, are those in which the density is
uniform over the sphere. [Note that for repulsive spin-independent
interactions with $SU(n)_{\rm spin}$ symmetry, the ground state
makes the density as uniform as possible.] It is convenient to
study these in the LLL in terms of their components $b_{m\mu}$,
$\mu=1$, \ldots, $n$, and $m=0$, \ldots, $N_v$. These components
form a matrix $B$, with $m$ labelling rows and $\mu$ labelling
columns. The density is uniform if and only if
\begin{equation}
BB^\dagger \propto I \label{isom}
\end{equation} where $I$ is the identity
matrix (that is, $B^\dagger$ is proportional to an isometry).
Solutions to this condition exist only when $0\leq N_v \leq n-1$.
For $N_v>n-1$, skyrmions will still occur, but the density will
not be completely uniform.

We now specialize to $n=3$ again, and consider the special cases
$N_v=1$ and $N_v=2$ for which uniform density condensates exist.
For $N_v=1$, $m=0$, $1$ only, so $B$ is a $2\times 3$ matrix. A
solution of eq.\ (\ref{isom}) for $B^\dagger$  is a complex number
times an isometry of ${\bf C}^2$ into ${\bf C}^3$. Thus this means
picking a two-dimensional subspace of the spin space ${\bf C}^3$.
Notice that $SU(2)_{\rm orb}$ acts by multiplication of $B$ on the
left, while $SU(3)_{\rm spin}$ acts by multiplication (by the
transpose) on the right, and the phase symmetry under $U(1)$ acts
on either side of $B$. Then the full space of solutions to eq.\
(\ref{isom}), for fixed mean particle number, is parametrized by
$SU(2)_{\rm orb}\times SU(3)_{\rm spin} \times U(1)/[SU(2)\times
U(1)]$, where the denominator represents the subgroup of
$SU(3)_{\rm spin}$ which has the same action as $SU(2)_{\rm
orb}\times U(1)$ on a particular solution $B$. This manifold is
equivalent to $SU(3)/{\mathbb Z}_3$. If we consider a quantum
state in which bosons condense in one condensate in this family,
then we can analyze it in terms of $N$, $\tilde{L}$,
$\tilde{L}_z$, and $SU(3)_{\rm spin}$ quantum numbers $(p,q)$. It
is easy to see that there can be no $SU(3)_{\rm spin}$ singlets
for $N_v=1$ as construction of such a singlet requires the use of
three distinct orbitals. Thus the broken-symmetry states cannot be
averaged over $SU(3)$ spin rotations to produce an $SU(3)$ singlet
(the closest one can get would be the BDC of Section~\ref{IV}, and
we believe that state should be interpreted in this way).
Nonetheless, we have shown the existence of configurations with single
units of vorticity. We have now analyzed the space of solutions
with uniform density using the full $SU(3)_{\rm spin}$ symmetry
group. For general $c_2\neq 0$, this symmetry is broken. In this
case, the form of the lowest-energy solution depends on the
energetics, and the manifold of lowest energy condensates (orbit
of the solution under the broken symmetries) is a submanifold of
that above with a lower dimension that depends on which solution
is chosen. Results of this analysis have been given in
Sec.~\ref{V}.

Similarly, for $N_v=2$, there are three orbitals, and the matrix
$B$ is now $3\times 3$. We see directly from eq.~(\ref{isom}) that
a solution for $B$ is proportional to a unitary matrix, and the
manifold of solutions is therefore $U(3)$. In terms of the
symmetries present for $c_2=0$ (i.e.\ using $SU(3)_{\rm spin}$
symmetry), this manifold arises as $SU(3)_{\rm spin}\times
SO(3)_{\rm orb}\times U(1)/SO(3)\times {\mathbb Z}_3$. In this
case the $SO(3)$ in the denominator, which is the unbroken
subgroup of the spin and orbital symmetries, has to be embedded
into $SU(3)$ as the group of $3\times 3$ orthogonal matrices. The
$U(1)$ transformations cannot be removed using the orbital or spin
symmetry groups.

The uniform-density skyrmions for $N_v=2$ and $c_2\neq0$ can be
found by using a careful choice of basis. For the ferromagnetic
case $c_2<0$, one expects that at each point on the sphere there
should be a nonzero (in fact, largest possible) spin density,
though its orientation varies over the sphere. If we use the basis
of single-particle $S_z$ eigenstates in the sequence
$\alpha=\dow$, $0$, $\up$, then one such solution is given by
$B\propto {\rm diag}(1,-1,1)$. In the particular solution given,
the spin density is $\dow$ at the north pole ($z=0$ in
stereographic coordinates), $\up$ at the south pole
($z\to\infty$), and the orientation elsewhere on the sphere can be
found by the relation of spin and orbital rotations as described
in the previous paragraph. In fact, the spin-density
$\langle\vec{S}\rangle$ itself wraps around the sphere, being
$(0,0,-1)$ at the north pole, $(0,0,1)$ at the south pole, and in
the $xy$ plane at the equator. This solution is the $\ell=1$
solution discussed in Sec.~\ref{V}. Other solutions are obtained
by multiplication of $B$ by an element of $SO(3)\times U(1)$. The
space of these solutions forms the manifold $SO(3)_{\rm
spin}\times SO(3)_{\rm orb}\times U(1)/SO(3)\simeq SO(3)\times
U(1)$, a submanifold of the full $U(3)$ we had before. The
solution $B={\rm diag}(0,-1,0)$ we began with here was chosen to
be invariant under the diagonal $SO(3)$ subgroup generated by
$\vec{\tilde{L}}+\vec{S}$. The other solutions, obtained by acting
with either $SO(3)_{\rm orb}$ or $SO(3)_{\rm spin}$, are invariant
under a similar subgroup with generators $\vec{\tilde{L}}$ plus a
(fixed) $SO(3)$ rotation of $\vec{S}$. These solutions are very
similar to the basic skyrmions for the spin-1/2 case, which appear
in the $\nu=1$ QH effect for electrons \cite{skkr}, even though 
here for spin 1 they have $N_v=2$. In fact those solutions 
for a condensate of
spin-1/2 bosons correspond to the $n=2$ component case with a
contact interaction and $SU(2)$ symmetry, as discussed briefly
above.

In the antiferromagnetic regime $\gamma>0$ for $N_v=2$, one
expects a different mean field ground state. As $\gamma \to 0^+$,
the ground state should approach a point in a submanifold of the
$U(3)$ manifold of uniform density states. Presumably this
submanifold is a different one from that for $\gamma<0$, where the
solutions always lie in the set we just described. In fact, we
expect that the ground states selected at $\gamma\to 0+$ are those
with the lowest spin density at each point on the sphere. At large
$\gamma$, the spin order at each point should take the polar form,
with vanishing spin density. This presumably cannot happen with a
uniform density. One expects the preceding uniform density
solutions to persist at small positive $\gamma$, and then
(possibly, above a nonzero critical value of $\gamma$) for the
density to become non-uniform. At large enough $\gamma$, we expect
the $N_v=2$ solution to contain two polar vortices, each with
vorticity one. These correspond to solutions found in Sec~\ref{V}.

For each of the manifolds that describe lowest-energy mean field
solutions, we can consider the configuration space of a point
moving on this manifold, and then quantize this motion (this is
known as semiclassical quantization of the collective
coordinates). This should reproduce the full space of states for
the $N\to\infty$ limit for these cases $N_v=1$, $2$, but we will
not go into these details here.

\end{appendix}

\end{document}